\documentclass[pdflatex,sn-mathphys]{sn-jnl}%
\usepackage{adjustbox}
\usepackage{mathtools}
\usepackage{graphicx} 
\usepackage{minitoc}
\usepackage{comment}
\usepackage{placeins}
\usepackage{setspace}

\let\Oldsection\section
\renewcommand{\section}{\FloatBarrier\Oldsection}

\let\Oldsubsection\subsection
\renewcommand{\subsection}{\FloatBarrier\Oldsubsection}

\jyear{2022}%
\theoremstyle{thmstyleone}%
\theoremstyle{thmstyletwo}%

\theoremstyle{thmstylethree}%

\newcommand{\detailtexcount}[1]{%
  \immediate\write18{texcount -merge -sum -q #1.tex output.bbl > #1.wcdetail }%
}

\begin{document}

\detailtexcount{sn-article}

\title[ ]{The interaction of transmission, mortality, and the economy: a retrospective analysis of the COVID-19 pandemic}

\author*[1]{\fnm{Christian} \sur{Morgenstern}}\email{c.morgenstern@imperial.ac.uk}

\author[1]{\fnm{Daniel J.}
\sur{Laydon}}

\author[1]{\fnm{Charles}
\sur{Whittaker}}

\author[2,3]{\fnm{Swapnil}
\sur{Mishra}}

\author[1]{\fnm{David}
\sur{Haw}}

\author*[1,3]{\fnm{Samir} \sur{Bhatt}}\email{s.bhatt@imperial.ac.uk}

\author*[1]{\fnm{Neil M.} \sur{Ferguson}}\email{neil.ferguson@imperial.ac.uk}

\affil[1]{\orgname{Imperial College London}, \country{UK}}
\affil[2]{\orgname{National University of Singapore}, \country{Singapore}}
\affil[3]{\orgname{University of Copenhagen}, \country{Denmark}}

\abstract{
The COVID-19 pandemic has caused over 6.4 million registered deaths to date and has had a profound impact on economic activity. Here, we study the interaction of transmission, mortality, and the economy during the SARS-CoV-2 pandemic from January 2020 to December 2022 across 25 European countries. We adopt a Bayesian Mixed Effects model with auto-regressive terms. We find that increases in disease transmission intensity decreases Gross domestic product (GDP) and increases daily excess deaths, with a longer lasting impact on excess deaths in comparison to GDP, which recovers more rapidly. Broadly, our results reinforce the intuitive phenomenon that significant economic activity arises from diverse person-to-person interactions. We report on the effectiveness of non-pharmaceutical interventions (NPIs) on transmission intensity, excess deaths, and changes in GDP, and resulting implications for policy makers. Our results highlight a complex cost-benefit trade off from individual NPIs. For example, banning international travel increases GDP and reduces excess deaths. We consider country random effects and their associations with excess changes in GDP and excess deaths. For example, more developed countries in Europe typically had more cautious approaches to the COVID-19 pandemic, prioritising healthcare, and excess deaths over economic performance. Long term economic impairments are not fully captured by our model, as well as long term disease effects (Long Covid). Our results highlight that the impact of disease on a country is complex and multifaceted, and simple heuristic conclusions to extract the best outcome from the economy and disease burden are challenging. 
}


\maketitle

\doparttoc %
\faketableofcontents %

\section{Introduction}\label{introduction}

As of August 2022, there have been over 580 million registered COVID-19 cases globally with over 6.4 million registered deaths \cite{who_dashboard}. However, these official counts substantially underestimate true COVID-19 disease burden \cite{Whittaker2021,Knutson2022}. The pandemic prompted the largest vaccination campaign in history \cite{Tatar2021-xq} with over 12.3 billion doses administered by August 2022 \cite{who_dashboard}. The pandemic had a profound impact globally due in part to the largely unprecedented scale of non-pharmaceutical interventions (NPIs) to control transmission. The effect of NPIs on SARS-CoV-2 community transmission has been studied in detail, for example \cite{liu2021,haug2020ranking,mendez2021systematic,sera2021cross,brauner_npi,sharma2021understanding,flaxman2020estimating,banholzer2022estimating}. Countries implemented NPIs, pharmaceutical interventions and economic support at different times and with differing individual specifications, allowing for a natural experiment to study their impact \cite{rosen2021pandemic}. It is of critical importance to understand how various suites of interventions affected individual countries' economic outlook and recovery.

The interaction between the pandemic and the economy have primarily been studied through two frameworks. The first framework integrates epidemiological models with economic models,  and stratifies the population, often by its types of economic activity (e.g. economic sector, work-from-home vs. in-person etc.), typically for a single country. Smith et al \cite{Smith_2009} is an early example of a Computable General Equvilibrium (CGE) model. The authors studied the interaction between school closures, vaccination and the economy during pandemic influenza in the UK in 2009, relative to a baseline scenario of no mitigations. These studies have focused less on estimating economic impacts of the pandemic, and more on exploring potential interactions between disease spread, policy interventions and the economy under a range of assumptions and hypotheses (e.g.\ \cite{deb2021,Eichenbaum2020,alvarez2021,acemoglu2021}). A minority of studies, e.g, \cite{haw2022optimizing}, estimate epidemiological parameters by fitting models to epidemic data. This approach is useful to study counterfactual scenarios and the impact of potential interventions, while {\it assuming} the dynamics that are imposed on the system. Integrated Epidemiological Economic models predate the COVID-19 pandemic, but have received increased focus due increased data availability, the number of countries and diversity of data. Several studies, such as Fenichel et al \cite{Fenichel_2011}, have focused on behavioural responses of individuals, and the models used in these studies tend to be highly mechanistic.

A second approach has been to estimate the interactions between transmission dynamics and the economy using statistical models, fitted to observational data, often from multiple countries. Longitudinal (panel) regression models have estimated the associations between economic variables, epidemiological variables and non-pharmaceutical restrictions. A variety of proxies have been used for economic variables, for example in \cite{godoy2022,verschuur2021,deb2021}. 

A small number of studies have used vector auto-regressive (VAR) models that incorporate temporal dynamics of the response variables, such as \cite{leibovici2022,jorda2022}. \cite{brodeur2021} provides a more comprehensive literature review on the economics of the COVID-19 crisis.

Here we develop a holistic framework which models the interaction between COVID-19 transmission, NPIs and economic outcomes. We consider four response variables: natural logarithm of 1 + excess deaths per 100,000 population (labelled $log ED$), weekly change in GDP (on a scale whereby a unit change represents a tenth of the change in GDP), weekly change in transit measured via Google mobility data (scaled such that a transit level of 1 corresponds to the baseline level prior to 6 February 2020), and the logarithm of $R_t$ (as a measure of disease transmission intensity). To jointly model the correlations within and between response variables we use a VAR model, which allows us to model temporal dynamics for each response variable. Within the VAR framework we consider NPIs as exogenous (fixed effects) variables, (we discuss this modelling choice in section \ref{exogeneity}). The NPI data we use is highly heterogeneous across countries and over time, and this allows us to study the efficacy of these measures across the pandemic, controlling for both vaccinations and SARS-CoV-2 variants. NPIs were imposed by policy makers to control disease spread, but we  also consider their broader behavioural and economic impact. The results reveal that each country needs to establish a balance of interventions given competing objectives and constraints they may face. We extend existing models to explore the dynamics of the above variables, in order to understand  different policy responses and interventions during the pandemic.

A key challenge of modelling the pandemic and the economy in tandem is that we observe data at different levels of aggregation, subject to differing lags in reporting. Epidemiological data is typically available at a daily or weekly frequency whereas most economic data is generally available quarterly or annually. We make use of a number of non-standard data sets and modelling techniques to overcome this problem. The GDP data we use infers the changes in GDP between official announcements and is provided by an OECD model \cite{oecdeco}, which matches official data closely. 

We focus on 25 European countries for which high quality weekly data are available. Our analysis spans 2020-2021 to generate a holistic picture of the pandemic's impact, which is not possible if considering only a single response variable (e.g. infection) or limited time period (e.g. the emergence and initial spread of SARS-CoV-2 in early 2020). We stop our analysis at the end of 2021 because other economic factors began to dominate in early 2022, most notably factors driving inflation and economic activity, which were independent of COVID-19.

\section{Results}\label{results}

\begin{figure}[htp]
\centering
\includegraphics[width=0.75\textwidth]{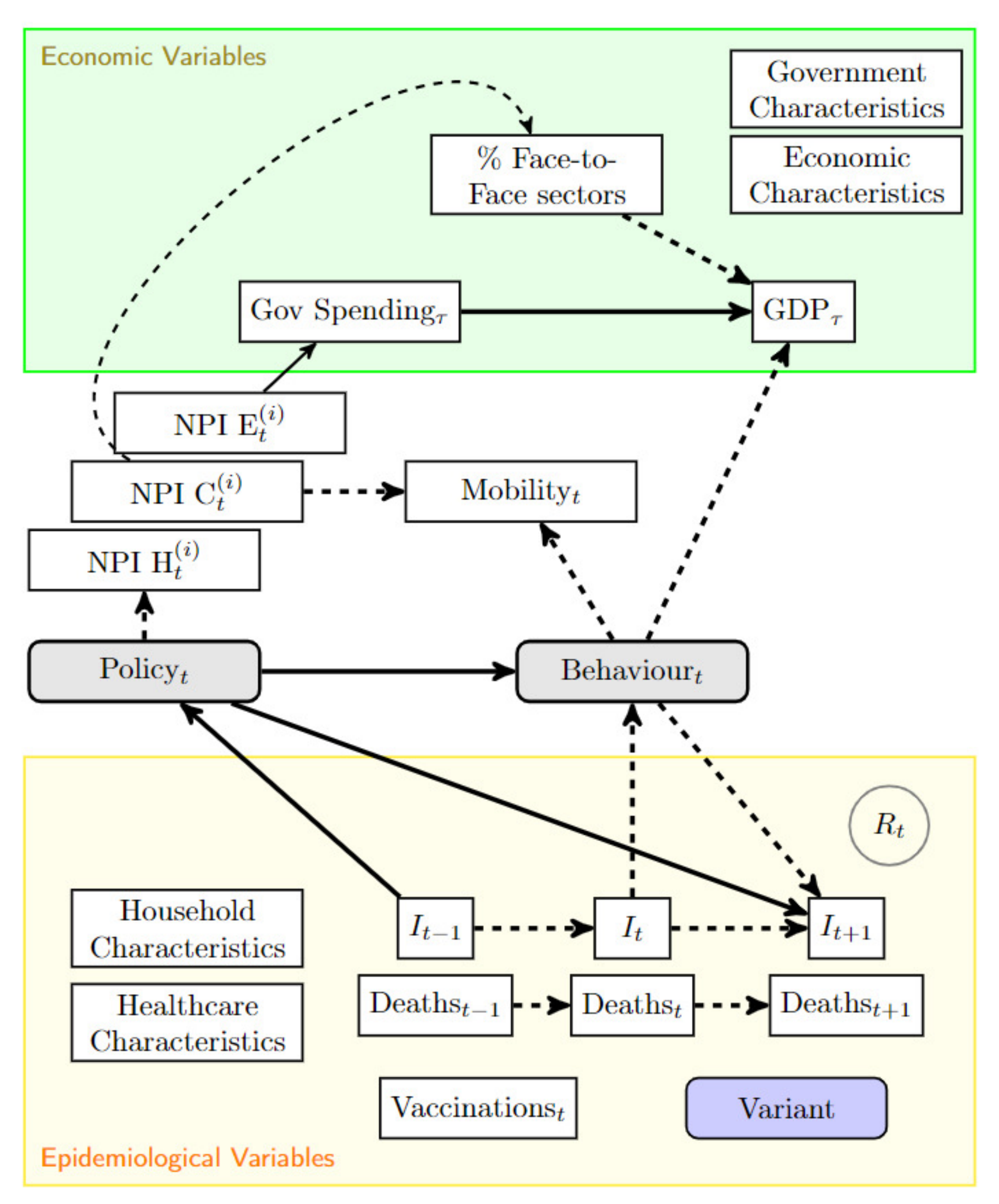}
\caption{Schematic model representation: Interaction of economic (top box) and epidemiological variables (bottom box). The middle section contains behavioural and policy variables. Variables in white square boxes are observed and available data in the model, purple boxes represent phylogenetic data available in the model and grey boxes represent latent variables. Solid lines are representing mechanistic links, dashed lines are hypotheses which we establish in the model. All variables are defined in Section  \ref{data}.}\label{model_schema}
\end{figure}

The conceptual framework of our study is summarised in Figure \ref{model_schema}. A detailed description of the data and the model are available in section \ref{methods}. A full list of parameter estimates is provided in Tables \ref{tab:VAR_coef} to \ref{tab:vaccination_coef} in the Supplementary Information. 

\begin{figure}[htp]
\centering
\includegraphics[width=1\textwidth]{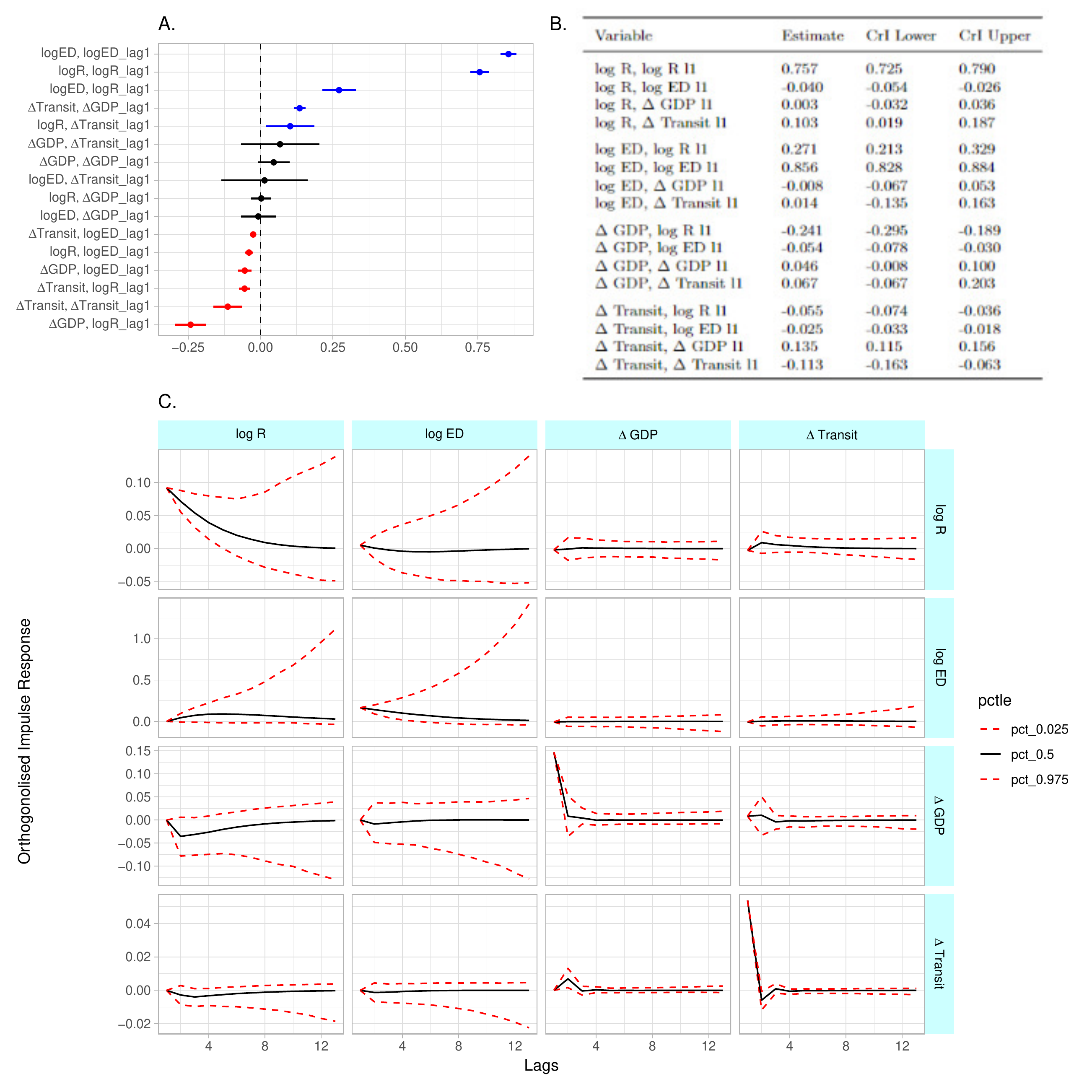}
\caption{Vector autoregression model estimates:  A. Estimated vector-autoregressive coefficients and 95\% credible intervals. Blue indicates  positive and significant coefficients, red shows negative and significant coefficients. B. Table of estimated vector-autoregressive coefficients. C. Orthogonalised impulse response functions. The columns are the variables to which the shock is applied, rows are the variables which we observe the impulse response for. Each plot shows mean estimates in solid black lines, with 95\% credible intervals in dotted red lines.
}\label{var_result_plot}
\end{figure}

We assess the temporal relationships between response variables firstly using regression coefficients of the VAR component of the model, and secondly using the impulse response functions, as summarised in Figure \ref{var_result_plot}. We provide the {\it Identification Strategy} in section \ref{identification}.

The most significant coefficients of the model were the auto-regressive coefficients of the response variables with themselves, lagged by one time unit (i.e. one week). The coefficients for $log ED$ and $log R$ were 0.856 (95\% CrI: 0.828-0.884) and 0.757 (95\% CrI: 0.725-0.790) respectively  (Figure \ref{var_result_plot} B), indicating generally slow changes in those variables. The autoregressive coefficient for $\Delta$ GDP was much lower (0.046, 95\% CrI: -0.008 to 0.100), indicating that GDP varies substantially over short time periods. For $\Delta$ Transit a negative autoregressive coefficient of -0.113 (95\% CrI: -0.163 to -0.063) indicates compensatory corrective behaviour.

We estimate a positive coefficient between $log ED$ and lagged $log R$ of 0.271 (95\% CrI: 0.213-0.329), representing, as expected, a positive correlation between deaths and transmission intensity. A positive coefficient between $log R$ and lagged $\Delta$ Transit of 0.103 (95\% CrI: 0.019-0.187) shows that mobility is associated with transmission intensity, as infections arise from previous person-to- person interaction. The positive coefficient between $\Delta$ Transit and lagged $\Delta$ GDP of 0.135 (95\% CrI: 0.115 to 0.156) suggests that individuals adjust their mobility behaviour as a function of past economic activity.

We estimate negative coefficients between $\Delta$ GDP and lagged $log R$ (-0.241, 95\% CrI: -0.295 to -0.189) and between $\Delta$ Transit and lagged $log R$ (-0.055, 95\% CrI: -0.074 to -0.036), indicating that transmission intensity depresses both GDP and mobility. 
Interestingly, we estimate small but significant negative coefficients between both $log R$ and lagged $log ED$ (-0.040, 95\% CrI: -0.054 to -0.026) and between $\Delta$ Transit and lagged $log ED$ (-0.025, 95\% CrI: -0.033 to -0.018), suggesting increases in excess deaths prompt some decrease in mobility and transmission, independently of interventions.

To analyse the effects of coefficients on the overall dynamics of the model, we evaluate impulse response functions  \cite{lutkepohl1990,pesaran1998generalized}. Impulse response functions represent the change over time  in three of the response variables due to a unit impulse (shock) in the fourth (Figure \ref{var_result_plot}C). We find that a positive impulse in $log R$ leads to a prolonged and significant increase in $log ED$, which peaks after 5 weeks and only decays slowly. A positive impulse in $log R$ also leads to a negative response in $\Delta$ GDP and a similar if smaller impact on $\Delta$ Transit, both of which peak after two weeks. Responses to impulses in the other response variables are consistent with coefficient estimates in Figure \ref{var_result_plot}A, but are relatively small in absolute terms, even when significant.

All impulses ultimately converge to zero but convergence is slowest for excess deaths, followed by $log R$ (as shown by the plots on the diagonal of Figure \ref{var_result_plot}C), due to these variables' high auto-regressive coefficient estimates. 

We observe heterogeneities in NPIs across  countries and time (Figure \ref{npi_plot}). In our VAR model, we include NPIs as predictor variables in two ways: the numerical level (stringency) of NPIs in each week, which we label as [level], and the change in that level from the previous week, which we label as [changes], (Tables \ref{OxCGRT_codebook_table_C},\ref{OxCGRT_codebook_table_E},\ref{OxCGRT_codebook_table_H}). We found that the inclusion of both the level and change in level improves model performance. We used Leave-One-Out cross-validation (LOO-CV) to estimate the pointwise out of sample prediction accuracy from our fitted Bayesian model \cite{vehtari2017practical}. The model with both levels and changes in NPIs outperformed a model with only levels of NPIs, with a expected log pointwise predictive density difference (ELPD-diff) of -80.9 (95\% CrI: -136.1 to -25.7). Our preferred model, and the model with only changes in level included (ELPD-diff of -23.8 (95\% CrI: -54.8 to 7.2)), are not statistically  different. We prefer the model with both changes and levels of NPIs as it allows us to explain the effect of NPIs more fully.

\begin{figure}[htp]
\centering
\includegraphics[width=1\textwidth]{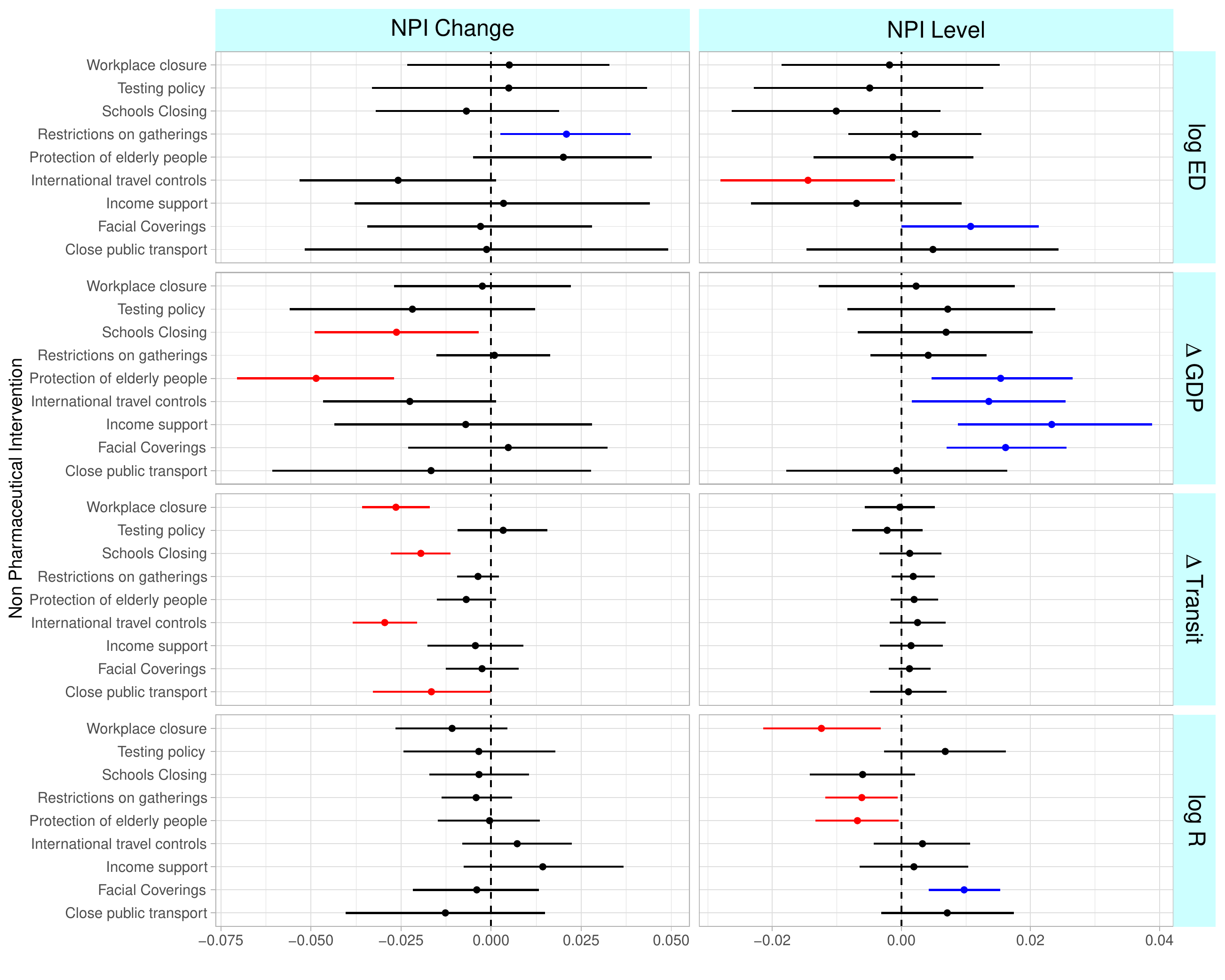}
\caption{Non-pharmaceutical intervention (NPI) effect sizes: Coefficient effect sizes (mean and 95\% credible intervals) from the regression group by response variable, NPI changes (left hand column) and NPI Levels (right hand column), NPIs are listed on the vertical axis. Blue indicates positive and significant coefficients and red indicates negative and significant coefficients.}\label{npi_result_plot}
\end{figure}

Figure \ref{npi_result_plot} shows estimated VAR model regression coefficients for NPI variables. Restrictions in workplace closure [changes] were statistically significant and associated with reductions in $\Delta$ Transit (-0.026, 95\% CrI: -0.036 to -0.017). Workplace closure restrictions [level] were associated with reduced $log R$ (-0.012, 95\% CrI: -0.021 to -0.003), indicating such closures led to a reduction of person to person contact rates. 

Restrictions on gatherings [level] were associated  with reductions in $log R$ (-0.006, 95\% CrI: -0.012 to -0.001), similarly to work place closures [level] (-0.012, 95\% CrI: -0.021 to -0.003). Restrictions on gatherings [changes] were associated with an increase in $log ED$ (0.021, 95\% CrI: 0.003-0.039).

International travel restrictions [changes] were associated with statistically significant reductions of $\Delta$ Transit (-0.030, 95\% CrI: -0.038 to -0.021), and with reductions [level] in $log ED$ (-0.014, 95\% CrI: -0.028 to -0.001). The association of $\Delta GDP$ and international travel restrictions [level] (0.014, 95\% CrI: 0.002 to 0.025), which also is significant for excess deaths [level] (-0.014, 95\% CrI: -0.028 to -0.001), is consistent with Clifford et al \cite{clifford2021strategies}, who discusses the effectiveness of different strategies of international travel restrictions, as those restrictions can slow down the importation of SARS-CoV-2 while allowing for a more open economy. 

School Closure [changes] and closure of public transport [changes] were both associated with reductions in $\Delta$ Transit. ``Protecting the elderly" [changes] (-0.049, 95\% CrI: -0.071 to -0.027) and ``Schools Closing" (-0.026, 95\% CrI: -0.049 to -0.003) [changes] NPIs were associated with reductions in $\Delta$ GDP. ``Protecting the elderly" [level] (-0.007, 95\% CrI: -0.013 to -0.000) was associated with a reduction in $log R$.

Several NPI levels had positive and significant associations with $\Delta$ GDP, in addition to International travel controls, Facial coverings [level] (0.016, 95\% CrI: 0.007 to 0.026), Protection of elderly people [level] (0.015, 95\% CrI: 0.005 to 0.027) and Income support [level] (0.023, 95\% CrI: 0.009 to 0.039). Each of these measures allow for a more open economy. In the case of Income support individuals were able to isolate, thus reducing transmission risk (as Income support itself is a transfer payment by governments and hence not adding to GDP).

Facial coverings [level] were associated with increases in both $log R$ (0.01, 95\% CrI: 0.004-0.015) and $log ED$ (0.011, 95\% CrI: 0.000-0.021). Facial coverings reduce the risk of infection, however as contacts increase we still see a net increase in infections and subsequent excess deaths. 

We included the dominant SARS-CoV-2 variant in our model (wild-type, Alpha, Delta and Omicron), together with the average number of vaccine doses received per capita, allowing the latter to vary by variant. Tables \ref{tab:Dominant_variant_coef} and \ref{tab:vaccination_coef} give estimates of the corresponding model parameters. 

The variant coefficient for $log R$ was significantly positive in the Delta and Omicron periods, and negative in the Alpha period. Stringent NPIs (e.g.\ gathering restrictions, workplace closures, school closures) were in place throughout most of Europe from December 2020 to May 2021, which likely accounts for the negative coefficients which we see for Alpha (Figures \ref{npi_plot} and \ref{dominant_variant}). Note we consider time periods over which there was a single dominant variant, and therefore differences between periods do not relect epidemiological characteristics of the variant itself. 

For $log ED$, the dominant variant had a significantly positive coefficient for the wild-type period, with smaller and non-significant coefficients for Alpha and Delta periods. The lower $log ED$ coefficient during the Alpha period is attributable to stringent NPIs, particularly high restrictions for isolating the elderly, and the initial vaccine roll-out. The elderly and those at higher risk of severe disease and death were prioritised, which likely affected both the Alpha and Delta periods. We estimate a large negative coefficient during the Omicron period, which is not quite significant at the 95\% level, probably because the Omicron period in the data was short. Variant is negatively associated with $\Delta$ GDP for the wild-type and Alpha, but positively associated for the Delta and Omicron periods, suggesting adaptation of the economy as COVID-19 restrictions eased. This adaption is a combination of easing of restrictions, businesses learning to operate during the pandemic with increased flexibility and time varying adherence to NPIs by individuals \cite{petherick2021worldwide}.

$\Delta$ Transit has positive coefficients for the Alpha period (0.013, 95\% CrI: 0.004-0.022), driving by the rebound in mobility during the first half of 2021, and negative coefficients during for Omicron (-0.050, 95\% CrI: -0.092 to -0.009) driven reduced mobility in the final months of 2021.

Vaccination was negatively associated with $log ED$ (-0.047, 95\% CrI: -0.092 to -0.002), indicating its primary role in reducing mortality, and $\Delta$ Transit (-0.02, 95\% CrI: -0.033 to -0.005). Vaccination was not significantly associated with $log R$ or $\Delta$ GDP. The lack of significant association with $\Delta$ GDP is because vaccination was only rolled out in 2021, and initially focused on the elderly and vulnerable, and not the working age population who make the largest contribution to economic activity, consistent with \cite{sonabend2021non} which studies this for the Delta variant in England.

\begin{figure}
\centering
   \includegraphics[width=1\linewidth]{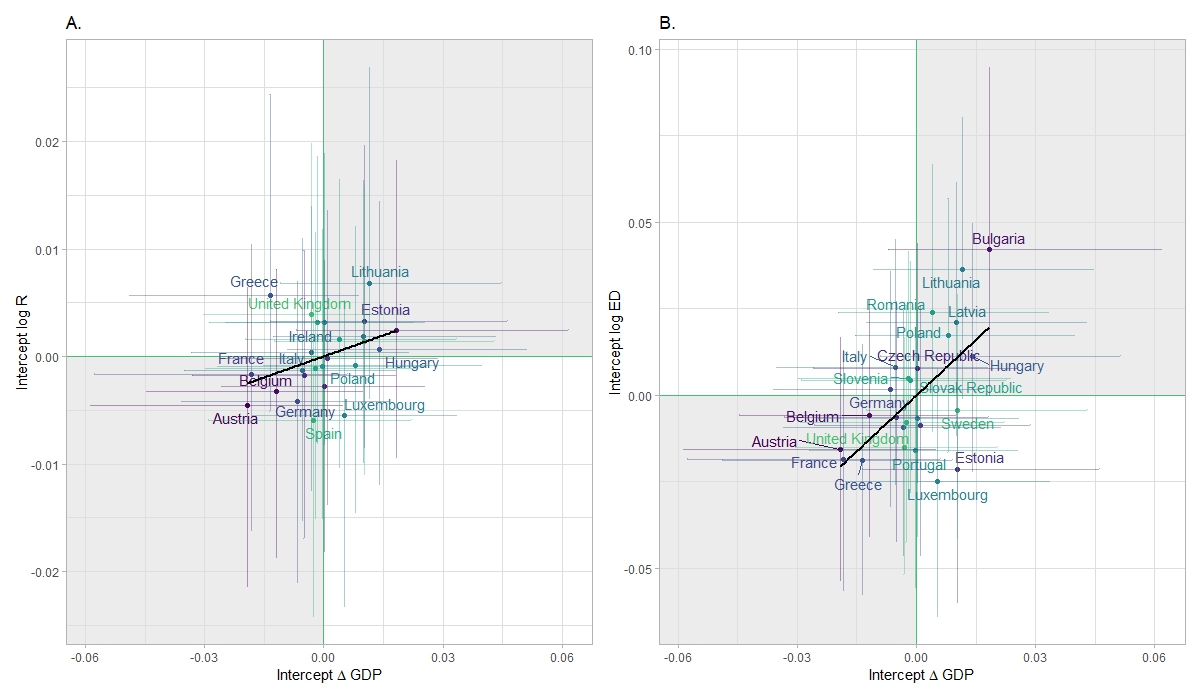}
   \includegraphics[width=1\linewidth]{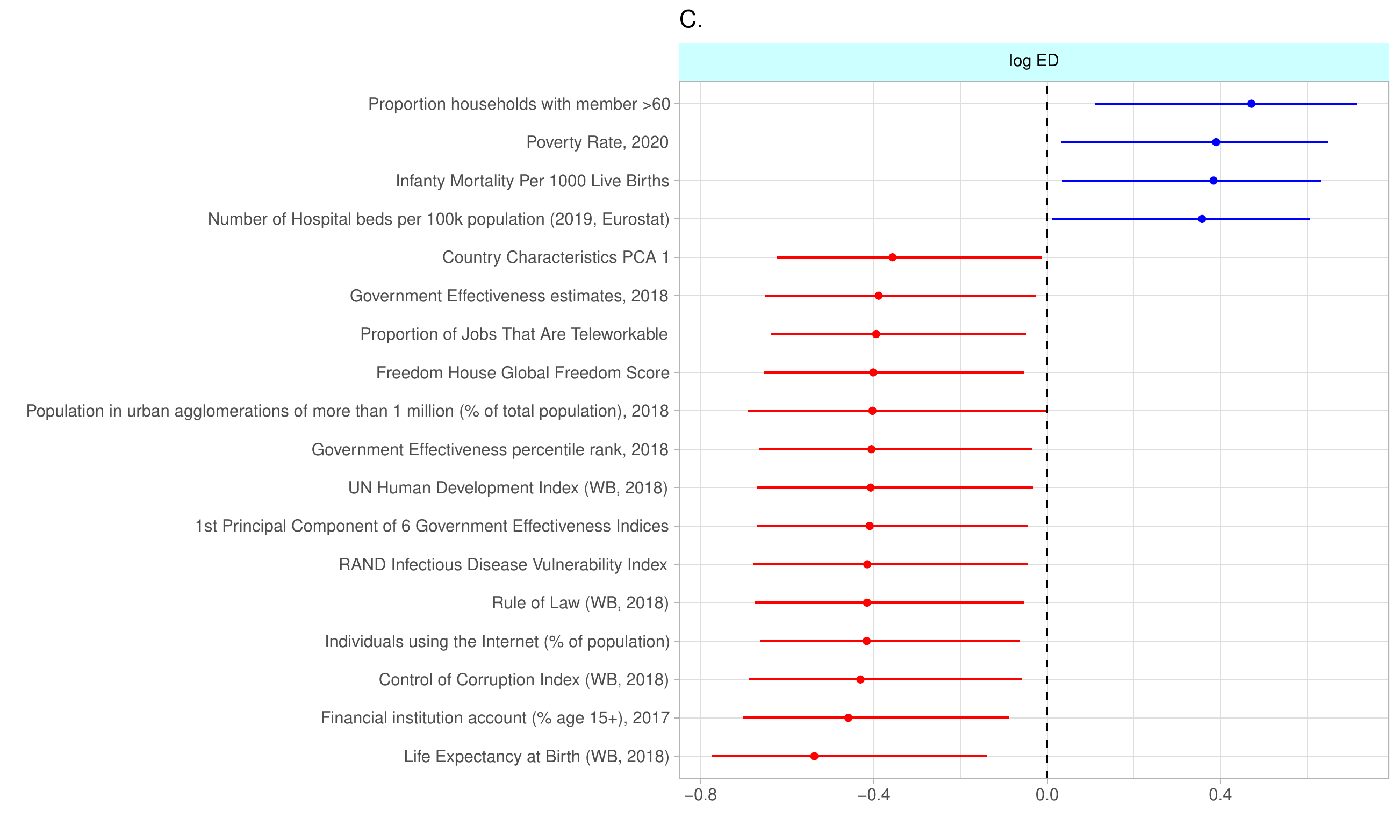}
\caption{(A.) Country specific intercepts for $log R$ vs $\Delta$ GDP.  (B.) Same as previous panel but for $log ED$ vs $\Delta$ GDP. (C.) Correlations of country specific intercepts with country specific characteristics for $log ED$ which are significant at the 95\% level (only significant characteristics at the 95\% level, all other data provided in SI \ref{cnt_spec_covariates}).}
\label{country_spec_figure}
\end{figure}

Countries had varied responses over the course of the pandemic. Eastern European countries imposed rigorous NPIs in March 2020 but less so from late 2020 onwards, compared with many Western European countries \ref{npi_eastern_europe_plot}. Our model included country specific intercepts as random effects  to represent residual variation after having taken account of all other ``known" factors (that is, a positive intercept for $log R$ in one country indicates a higher than average $log R$ over the time period).  Figure \ref{random_effects_plot} shows the variation across countries for all response variables. There is only minor unexplained variability in $\Delta$ Transit compared with the other three response variables. The ordering of countries for excess deaths and transmission intensity are broadly in line, however there is some change in the rank-order due to differences in healthcare systems. 

We find a positive correlation between $\Delta$ GDP and $log ED$ country intercepts (0.23, 95\% CrI: -0.21 to 0.59 (Figure \ref{country_spec_figure}B), suggesting some trade-off between pandemic mortality and GDP that is independent of interventions (i.e. factors like behavioural changes). $\Delta$ GDP and $log R$ country intercepts also have positive correlation (0.09, 95\% CrI: -0.33 to 0.46, see Figure \ref{country_spec_figure}A). 

We consider the effect of country-specific intercept parameters on country characteristics. We include several characteristics across Government, economic, Household and Healthcare, as listed in Table \ref{characteristics_table}. Resulting correlations are summarised in Figure \ref{country_spec_figure}C and listed in full in Table \ref{covariates_result_table}. For $log ED$, we find positive correlations for the proportion of households with members over 60, Poverty rates, and Infant mortality per 1000 live births. Similarly we find negative correlations between the $log ED$ intercept and: Life expectancy at birth; access to financial institution accounts; Control of corruption; Individuals using the internet; Rule of Law; Infectious Disease vulnerabilty; the first principal component of 6 Government effectiveness indices; Human Develepmeny Index; Government effectiveness percentile rank; proportion of jobs that are teleworkable, and Freedom House global freedom score. 

The $\Delta GDP$ intercept is negatively correlated with: Health expenditure as a fraction of GDP; life expectancy at birth; and Proportion of the population Living within 60min of an Urban Centre (Figure \ref{covariances_cor_full_1}) at the 80\% CrI. Therefore, countries with higher health expenditure had lower changes in GDP. The country-specific intercepts for $\Delta Transit$ and $log R$ were not significantly correlated with any country-level characteristics we examined (Figure \ref{covariances_cor_full_2})

We consider different approaches to dimensional reduction and clustering of the country specific characteristics in the supplementary information \ref{PCA_section}.

We are interested in the causal relationships of our response variables. Clearly, a cause cannot precede its the effect. If a variable $x$ affects another variable $y$, the former should help improve the predictions of the latter. We again use LOO-CV to estimate the pointwise out-of-sample prediction accuracy from our fitted model. We systematically remove predictive variables from the model: if variable x is removed, it is removed from all equations, except the equation of x on itself (e.g. removing $log R$ removes it from the right hand side of the equations of $log ED$, $\Delta$ GDP and $\Delta$ Transit, but not the right hand side of $log R$ itself). We do this for each response variable individually, all combinations of two response variables and removing all response variables (in which case each variable uses only itself). The results are summarised in Table \ref{tab:LOO_CV}, ordered by decreasing expected log pointwise predictive density (ELPD) difference.

The full model containing all variables has the best ELPD, although models excluding $\Delta$ Transit, $\Delta$ GDP or both of these are not statistically different at the 95\% level from the full model. Excluding $log R$ or $log ED$, and any combination containing these predictors, the difference is statistically significant. $log R$ is the most important predictive variable, as it leads to the largest negative ELPD differences, either on its own (compared to other single variable exclusions) or in combination with other variables excluded. Excluding all variables, so that each response variable is a function of only its past observations, leads to the largest ELPD negative difference indicating the worst performing model with respect to sample prediction accuracy. Together with the above vector-autoregressive coefficient estimates and the impulse response function, we have further evidence that transmission intensity increases excess deaths and negatively changes GDP.

\begin{table}
\begin{center}
\begin{tabular}[t]{lrrrr}
\toprule
Excluded Variable & ELPD-diff & SE-diff & CrI Lower Bound & CrI Upper Bound\\
\midrule
Full Model & 0.000 & 0.000 & 0.000 & 0.000\\
$\Delta$ Transit & -5.690 & 4.685 & -15.060 & 3.681\\
$\Delta$ GDP & -9.984 & 5.896 & -21.777 & 1.808\\
$\Delta$ GDP \& $\Delta$ Transit & -14.264 & 7.387 & -29.039 & 0.511\\
$log ED$ & -40.741 & 10.350 & -61.440 & -20.042\\
$log ED$ \& $\Delta$ Transit & -48.823 & 11.441 & -71.706 & -25.941\\
$\Delta$ GDP \& $log ED$ & -51.353 & 12.469 & -76.291 & -26.415\\
$log R$ & -69.044 & 13.484 & -96.012 & -42.076\\
$log R$ \& $\Delta$ Transit & -73.989 & 14.424 & -102.837 & -45.140\\
$\Delta$ GDP \& $log R$ & -84.924 & 14.819 & -114.563 & -55.286\\
$log R$ \& $log ED$ & -117.163 & 18.069 & -153.301 & -81.026\\
All Variables & -145.501 & 20.373 & -186.248 & -104.754\\
\bottomrule
\end{tabular}
\caption{Leave-One-Out cross-validation (LOO-CV) to estimate the pointwise out of sample prediction accuracy of fitted Bayesian models. The full model is that described in the main text, excluded variables are removed from the prediction variables of the model (except for variables on themselves). Expected log pointwise predictive density (ELPD) differences and Standard Errors are reported.}\label{tab:LOO_CV}
\end{center}
\end{table}

To further assess the predictive performance of our model, we compare the 1-timestep ahead forecast from our model to the naive forecast, which assumes the next value of a response variable is the same as its current value, that is $\hat{x}(t+1)=x(t)$. We find that our model improves predictions of all response variables, with a reduction of root mean squared error of 8\% for transmission intensity, 6\% for excess deaths, 27\% for $\Delta$ GDP and 32\% for $\Delta$ Transit (Figure \ref{forecast_plot}). The improvements are due to improved forecasts in the tail of the distributions as opposed to the centre. This is expected, as the centre represents relatively stationary time periods. The model however does not have any awareness of e.g. holiday effects or other exogenous dynamics, and hence does not outperform the naive forecast during these periods. In the tails, i.e. during periods of high transmission intensity, the model performs better than the naive forecast.

\section{Discussion}\label{discussion}

The interplay between transmission intensity, excess deaths, population mobility and GDP are highly complex, and cannot  be considered in isolation. Here, we study their interaction holistically, for 25 European countries, during the first two years of the SARS-CoV-2 pandemic. Higher transmission intensity decreased GDP and increased excess deaths. Intensified public health interventions reduced mobility, transmission intensity, excess deaths and GDP. However, as our estimates allow for asymmetric relationships, changes in GDP alone (independent of control measures) were seen to have no significant impact on either transmission intensity or excess deaths. These findings generalise and extend results from previous studies \cite{leibovici2022}, and make them more robust. Broadly, our results reinforce the intuitive phenomenon that significant economic activity arises from person-to-person interaction. Reductions in such interactions - by governmental mandate or behavioural change - reduce transmission, but also harm economic activity.

NPIs affect transmission, GDP and excess deaths in a variety of ways, but often act differently to each other. International travel restrictions reduce excess deaths, but increase economic activity, as more domestic activity can take place where importation of the virus is slowed due to quarantine, testing or outright travel bans \cite{clifford2021strategies}. As found previously \cite{sharma2021understanding}, workplace closures are associated with reductions in transmission intensity. As highlighted further below, we also find that specific economic characteristics of each country reduce transmission intensity. For example, countries with more face-to-face service sectors have higher transmission intensity (see Table \ref{covariates_result_table}). 

Facial coverings (masks of all types) are an important intervention, as their societal costs and impacts are relatively small and no future economic costs are associated with facial coverings \cite{leech2022mask}. Economically, facial coverings allow for greatly more open societies, and therefore economic activity. The impact of facial coverings on transmission intensity and excess deaths is more difficult to assess, as this is a function of adherence and face covering quality. We see positive associations between facial coverings (or mask mandates), and both excess deaths and transmission intensity. This may seem counter-intuitive, but can be regarded as a net effect of re-opening. While face masks allow increased social contact, and protect against transmission (albeit imperfectly), their adoption, concurrent with the easing of social distancing that they allow, can increase transmission and therefore excess deaths. Here, we do not assess the adherence to mask wearing by individuals, and only consider the policy. 

Income support is an important intervention but direct interpretation of this coefficient requires caution. Income support itself is a government transfer payment which has no impact on GDP. The positive coefficient which we observe in the model with respect to $\Delta$ GDP is the indirect result from two effects. First, income support enables individuals to continue to consume, even if they cannot work. Second, it enables businesses to continue to employ staff and maintain basic functions through the most significant restrictions, so that they can restart activity as soon as possible.   

The behavioural response of individuals is an important component. Fenichel et al \cite{Fenichel_2011} used a highly mechanistic model to estimated the impact of behaviour on person-to-person contact and the estimates of $R_t$. The most direct proxy for behaviour in our model is transit mobility and, according to our estimates, individuals reduce their transit mobility as disease transmission and excess deaths increase, exhibiting risk averse behaviour, and individuals increase their transit mobility in response to increasing economic activity, exhibiting risk seeking behaviour. Further, individuals reduce their transit mobility in response to NPIs (in the previous time period). However, in the longer term, the level of restrictions has no statistically significant impact.

Country-specific intercepts in our model account for residual variation, after considering the past dynamics of the response variables, NPIs, vaccination and the dominant variant of SARS-CoV-2. Country-specific intercepts for excess deaths and GDP were correlated with a number of country-specific characteristics [Figure \ref{country_spec_figure}], suggesting that more developed European countries had more cautious approaches to the pandemic, prioritising healthcare and lower mortality over economic performance. This can also be seen in Figure \ref{random_effects_plot_2}, where we can see the positive slope between country-specific intercepts for excess deaths and changes in GDP. Although country-specific intercepts account for latent factors, the collinearity between the trends in GDP and excess deaths suggests a trade-off. Additionally, the choice to alter economic performance is not binary, and need to carefully calibrate according to a country's economic outlook, policy objectives and fiscal ability to sustain restrictions and associated support.

A number of country characteristics are associated with lower excess deaths (Figure \ref{country_spec_figure}). Most characteristics reflect the general development level of a country, such as the Global freedom score, Control of Corruption, Internet usage, and the proportion of jobs which are teleworkable. 

We find, counterintuitively, but consistently with previous studies \cite{haw2022optimizing}, that larger hospital capacity is associated with greater excess deaths. This association arises because a higher healthcare capacity allows policy makers to delay introduction of more intensive, but ultimately necessary, NPIs for longer than smaller capacity would allow, resulting in more infections and deaths. We also note that the capacity and the quality of a healthcare system are not necessarily equivalent, and vary across countries.

A limitation of our analysis is that we do not consider long-term health and economic impacts of the pandemic, such as those caused by loss of schooling \cite{school_closure_oecd}, mental health impacts \cite{joseph2022physical} and long COVID \cite{long_covid}, as our work is focused on the short term impact of the pandemic over the first 2 years. Another limitation is our assumption of a linear relationship between the four response variables and other covariates (such as NPIs). More generally, we do not model any latent processes in our analysis - e.g. underlying infection incidence - and we are therefore limited in our ability to mechanistically represent relationships between response variables and interventions. Conversely, our omission of latent processes has the advantage that our model is less constrained by structural assumptions than a more mechanistic modelling framework.

Future research should address the above limitations. In particular, despite the greater structural constraints, more mechanistic representations of both disease epidemiology (e.g. extending the framework of Brauner et al \cite{brauner_npi} to include economic impacts) and the economy (e.g. by embedding a macroeconomic models into the modelling framework) should provide greater insight into the relationships between government policies, country differences and the resulting health and economic effects of the pandemic.

\section{Methods}\label{methods}

We consider the time period from 1 January 2020 to 31 December 2021 across 25 European countries. We model the dynamics of four response variables: i) excess deaths ($log ED$), ii) changes in GDP ($\Delta GDP$), iii) changes in mobility ($\Delta Transit$) and iv) transmission intensity ($log R$). We further model how each response variable is affected by different NPIs, given vaccination rates and the dominant SARS-CoV-2 variant. Figure \ref{model_schema} provides an overview schematic of our model. Data that are available at daily frequency are indexed by $t$, and data that are available at lower frequency is index by $\tau$. We down-sample all variables to weekly frequency in the model.
$log$ is the {\it natural} $log$ throughout the text.

\subsection{Data}\label{data}

During the COVID-19 pandemic, epidemiological data availability has been generally good, with most available at daily or weekly frequency. Economic data which has seen the most improvement during the pandemic, with several attempts made to derive proxies of economic activity that are updated more often (daily or weekly) than traditional economic indicators. Examples include electricity consumption (\cite{chen2020}) and shipping nowcasts (\cite{verschuur2021,cerdeiro2020}). These proxies have been used to generate weekly GDP Nowcasts \cite{oecdeco}, and constitute a good compromise between frequency and their ability to measure economic impact directly and accurately. However, significant data needs remain with respect to economic data to model the epidemiological-economic interaction more accurately \cite{haw2022data}.

We use Stan for parameter inference with Hamiltonian Monte Carlo. To ensure that sampling is efficient we scale variables to be centred around zero and of similar magnitude, with scaling coefficients chosen accordingly. We also note that all the response variables are changes as $R_t$ is a rate and excess deaths are deviations from the expected number of deaths in a given time period.

{\bf GDP Nowcasts} The OECD GDP Tracker \cite{oecdeco} provides weekly GDP Nowcasts for 46 countries (OECD \& G20). A panel of Google Trends data across 215 categories and 33 topics, which are relevant to economic activity, is used to construct a 2-step model to provide nowcasts. In Figure \ref{oecd_nowcast_validation}, we show model validation against UK GDP data (available monthly rather than quarterly). A full set of validations is available on the OECD website. Figure \ref{response_variables_plot_GDP} plots the GDP nowcast data across countries. We define one additional measure, the pre-pandemic trend growth $g_c$, as the mean rate of GDP growth between 2016 and 2019 for each country. We adjust the GDP nowcast data to remove this pre-pandemic growth trend in order to adjust for inherent differences in European economies' growth . Equivalently, this means that we evaluate GDP impacts of the pandemic relative to a no-pandemic counterfactual scenario. In Figure \ref{gdp_impact}, we show overall GDP growth over the two year period, with and without removal of the pre-pandemic growth trend. We define a response variable $\Delta GDP$ in our model, representing the weekly change in GDP: $\Delta GDP_{c,t} = (GDP_{c,t}-GDP_{c,t-1}-g_c)/10$, where $g_c$ is the weekly pre-pandemic growth, $t$ is week, and $c$ indexes the country. We scale by a factor of $\frac{1}{10}$ to ensure that the dimensions of $\Delta GDP$ are producing parameter estimates in the same order of magnitude as other coefficients to ensure that the estimation converges more quickly.

{\bf Time-varying Reproduction Number $R_t$} To assess temporal variations in SARS-CoV-2 transmission, we use the package EpiNow2 based on \cite{abbott2020}. EpiNow2 uses case and death notification data to generate daily estimates of time-varying reproduction numbers over a 12-week rolling window, which is available for all our 25 European countries. A key criterion for the choice of model for $R_t$ estimates was the ability to provide consistent estimates across countries. We have evaluated the EpiNow2 estimates against other models such as \cite{arroyo-marioli2021,sangeeta_rt_est}. In Figure \ref{rt_comparison_plot}, we provide a visual comparison that broadly  indicates models provide similar $R_t$ estimates. We use $log R$ as the response variable (Figure \ref{response_variables_plot_logR}), which is the transmission intensity.

{\bf Mobility Data} Google mobility data \cite{google_mobility} provides a range of measures of human mobility. We consider {\it Transit} as this captures the willingness of individuals to use public/mass transit networks, both to commute to work and to otherwise leave their primary residence and participate in economic activity. Transit is also of interest for SARS-CoV-2 transmission, as it captures potential contacts outside individuals' households, unlike other modes of transport which may not involve such contacts (e.g. car travel). Transit data is available at daily resolution, for each country of interest, and is represented as a percentage change from the median value, for the corresponding weekday, during the period of 3 January to 6 February 2020. In the model we use $\Delta Transit_t=(Transit_t - Transit_{t-1})/100$, where $Transit_t$ is a 1 week moving average to capture the behaviour over the full period (Figure \ref{response_variables_plot_transit}). 

{\bf Excess deaths Data} We use daily excess deaths estimates generated by the Economist \cite{economist_excess_deaths}. We use excess deaths rather than laboratory-confirmed COVID-19 deaths because countries varied widely in both their testing and their definition of COVID-19 deaths \cite{economist_excess_deaths,watson2021leveraging}. We use $log(excess\_deaths\_per\_100k+1)$ as the response variable (Figure \ref{response_variables_plot_logED}), which varies between -2 and 2.
We account for the time between  COVID-19 infection and death by leading the excess deaths response variable by 3 weeks, in order to align with other response variables (assuming a 5-day mean incubation period and mean 15-day delay from symptom onset-to-death \cite{sharma2021understanding}).

{\bf Non-pharmaceutical Interventions:} We obtain NPI data from the Oxford Blavatnik School of Government Covid-19 Government Response Tracker (OxCGRT) \cite{Oxford_BSG}. OxCGRT provides systematic data on interventions implemented by governments on a daily basis, since the start of the pandemic. The policy actions are split into 3 categories: 8 containment and closure policy indicators (C1-C8), 4 economic policy indicators (E1-E4) and 8 health system policy indicators (H1-H8). For each measure a score is available that reflects the severity / scale of the restriction, see Tables \ref{OxCGRT_codebook_table_C}, \ref{OxCGRT_codebook_table_E}, \ref{OxCGRT_codebook_table_H}. 
We use both the level of each NPI and the first difference ($x_t - x_{t-1}$ for NPI $x$ at time $t$) as explanatory variables.
Figure \ref{npi_plot} plots all NPIs considered in our analysis throughout the time period. 

{\bf Covid-19 Variant Data} SARS-CoV-2 variant data is sourced from \cite{covariant}, which provides a summary view by country and week based on GISAID data. We utilise the Nextstrain Clade, Pango Lineage and WHO Label mapping to map all Nextstrain Clades into WHO Labels, for use in the model. For each week we pick the dominant strain in each country as the strain with the majority of sequenced samples (Figure \ref{dominant_variant}). We observe a data gap for Hungary, which is missing weeks 9-45 in 2021. We estimate data for Hungary during this time period by using the average of the data from surrounding countries, weighted by border length.

{\bf Covid-19 Vaccination Data} Our World In Data \cite{mathieu2021global} provides a range of vaccination statistics. We use the total number of vaccine doses administered over time. dividing by the total population \cite{eurostat_data} to calculate the number of doses delivered per capita (data plotted in Figure \ref{vaccination_plot}). Unfortunately there is no consistent data across countries with a breakdown of vaccinations by age group (or any other characteristic).

{\bf Country Characteristics Data} We summarise the sources for the country characteristics data in Table \ref{characteristics_table}.   

\begin{table}[ht]
\begin{tabular}{@{}lrrrr@{}}
\toprule
Data Item & Reference & Number of & Data & Reference\\
 & Year & Countries & Source & \\
\midrule
\multicolumn{4}{l}{\textbf{economic Data}}\\
\hspace{2em} \% Face-to-Face sectors & 2020 & 17 & EuroStat & \cite{eurostat_data}\\
\hspace{2em} Labour Markets & 2018-20 & 24 & WB & \cite{wb_data,lakner2020much,dingel2020many} \\
\hspace{2em} Health Expenditure & 2016 & 24 & WB & \cite{wb_data}\\
\hspace{2em} Electric Power Consumption & 2014-18 & 25 & WB & \cite{wb_data} \\
\hspace{2em} Net Forest Depletion \% of GNI & 2018 & 24 & WB & \cite{wb_data} \\
\multicolumn{4}{l}{\textbf{Heath Systems Data}}\\
\hspace{2em} \# Physicians per 100k pop & 2019 & 17 & EuroStat & \cite{eurostat_data}\\
\hspace{2em} \# Hospital beds per 100k pop & 2019 & 24 & EuroStat & \cite{eurostat_data}\\
\hspace{2em} Mortality Statistics & 2018 & 24 & WB & \cite{wb_data} \\
\hspace{2em} RAND Inf. Disease Vulnerability & 2016 & 25 & RAND & \cite{rand} \\
\multicolumn{4}{l}{\textbf{Household Data}}\\
\hspace{2em} Household Composition & 2018 & 23 & UN & \cite{un_data}\\
\hspace{2em} Financial Inclusion & 2018 & 25 & WB &\cite{wb_data} \\
\hspace{2em} Cultural Factors (e.g. Obedience) & 2019 & 25 & Paper & \cite{schulz2019church} \\
\multicolumn{4}{l}{\textbf{Government Data}}\\
\hspace{2em} Government Effectiveness & 2018 & 24 & WB & \cite{wb_data}\\
\hspace{2em} \% Seats Women in Parliament & 2018 & 24 & WB & \cite{wb_data} \\
\hspace{2em} Property Rights & 2018 & 24 & HF & \cite{heritage_data}\\
\hspace{2em} Labour, Regulatory Freedom & 2018 & 24 & HF & \cite{heritage_data} \\
\hspace{2em} Global Freedom Score & 2021 & 25 & FH & \cite{freedomhouse_data} \\
\hspace{2em} Rule of Law & 2018 & 24 & WB & \cite{wb_data} \\
\hspace{2em} Control of Corruption & 2018 & 24 & WB & \cite{wb_data} \\
\multicolumn{4}{l}{\textbf{Societal Data}}\\
\hspace{2em} Population Density & 2017/18 & 25 & WB & \cite{wb_data} \\
\hspace{2em} Poverty Statistics & 2018-20 & 24 & WB & \cite{wb_data,lakner2020much} \\
\hspace{2em} GINI Income Inequality & 2018 & 24 & WB & \cite{wb_data} \\
\hspace{2em} UN Human Development Index & 2018 & 24 & WB & \cite{wb_data} \\
\hspace{2em} Prevalence of Undernourishment & 2018 & 24 & WB & \cite{wb_data} \\
\botrule
\end{tabular}
\caption{Country Characteristics: Overview of data items, reference year and coverage of countries. Abbreviations for data sources are Heritage Foundation (HF), World Bank (WB), International Monetary Fund (IMF), United Nations (UN), Freedom House (FH) }\label{characteristics_table}%
\end{table}

\subsection{Model}\label{model}

We consider a panel vector auto-regressive (VAR) model of the response variables to capture their temporal dynamics. Both NPIs, and changes in NPIs, are considered as exogenous variables. Other time-varying predictors examined are cumulative vaccination doses administered per capita, and the dominant variant of SARS-CoV-2. 

We express the model as a Bayesian multi-level mixed-effects model, defined by Equations 1-5. The model is partially pooled as we assumed predictor coefficients are common to all countries while also estimating country specific intercepts, $\mu_c$, where countries are indexed by $c$. The effect is to share information between countries, while still permitting between-country variation. 

\begin{eqnarray}
Y_{t,c} &\sim& {\tt MVN} \big(y_{t,c},\Sigma_{u_{t,c}}\big)\\
\nonumber y_{t,c} &=& {\bf \mu_c} + \sum_{k=1}^p \Phi_k y_{t-k,c} + {\bf \lambda} \cdot x_{t,c} + {\bf \delta} \cdot \Delta x_{t,c} + \\
\nonumber && {\bf \nu} \cdot vacc_{t,c}  + {\bf \psi} \cdot \sum_{\substack{j\in\{WT,Alpha,\\ Delta,Omicron\}}} \Psi_{j,t,c}  \\
\nonumber where:\\
\nonumber y_t &=& \big( y_{1t},y_{2t},...,y_{Nt}\big)^{\top}\\
&=& \big( {\tt \log R}_t, \log{\tt Excess\>Deaths}_t, \Delta {\tt GDP}_t,\Delta {\tt Transit}_t, \big)^{\top}\\
x_t &=& {\tt NPI}_{t-1} \\
vacc_t &=& {\tt Average \>Vaccinations\>per \>person }_t
\end{eqnarray}

$y_{t,c}$ is a vector of the values of the response variables (as defined in Eq 3) for time $t$ and country $c$. $\Psi_{j,t,c}$ is a variable for each dominant variant of SARS-CoV-2 in a country $c$ at time $t$. $\Psi_{WT,t,c}$ is each to 1 always (and acts the same as an intercept term), $\Psi_{Alpha,t,c}$ is zero if at time $t$ in country $c$ WT is the dominant variant but 1 otherwise, $\Psi_{Delta,t,c}$ is zero if at time $t$ in country $c$ WT or Alpha are the dominant variants but 1 otherwise and $\Psi_{Omicron,t,c}$ is 1 if Omicron is the dominant variant at time $t$ in country $c$ and zero otherwise. $u_{t,c}$ is the error term at time $t$ in country $c$. $\Phi_k$ is a 4 by 4 matrix of coefficients to be estimated, and $\lambda_i,\delta_i,\nu_,\psi_i$ are each vectors of length 4 of coefficients to be estimated, where $i$ is the index going over each one of the factors for NPIs, changes in NPIs, dominant variant and vaccination respectively.

We impose the following priors on the coefficients. The priors for the coefficients of each response variable in the VAR model are normally distributed with zero mean and variance $\tau$. Similarly, we impose a normal distribution with zero mean and variance $\tau$ on the coefficients of the NPI variables, vaccine doses per capita and dominant variant. The parameter $\tau$ is manually set to 1, corresponding to a relatively uninformative prior. The country specific intercepts have a normal prior with zero mean and variance $\sigma^2$, where $\sigma$ has a Cauchy(0,2) prior.

Group-level parameters are assumed to come from a multivariate normal with an unknown covariance matrix, which we also assign a prior to. $\Sigma_u$ is decomposed into a vector of variances $(\sigma_1,...,\sigma_N)$ and a correlation matrix $\Omega_N$, which is given an LKJ prior (spherical prior on correlation matrices, \cite{burkner2018a}) with parameter $\xi=2$.  

\begin{eqnarray}
\Phi_{k,i,j} &\sim& N(0,\tau^2) \>\>\> \forall k=1,i\in[1,4],j\in[1,4]\\
\lambda_i,\delta_i, \nu_i, \psi_i &\sim& N(0,\tau^2) \>\>\>\forall i\\
\mu_{c}&\sim& N(0,\sigma^2) \>\>\>\\
\sigma &\sim& cauchy(0,2)\\
\Omega_N &\sim& {\tt LKJ}(\xi)
\end{eqnarray}

\begin{table}[ht]
\begin{tabular}{@{}lrrr@{}}
\toprule
Parameter & Type &  Value & Range\\
\midrule
p & VAR & 1 & \\
panel\_id & VAR & Country & Leave one out sensitivity\\
$\tau$ & prior & 1  & (0.2,2) \\
$\sigma_i$ & prior & cauchy(0,2) & \\
$\xi$ & prior & 2 & (1,2,4)\\
adapt\_delta & Stan & 0.9 &  \\
step\_size & Stan & 0.01 & \\
max\_treedepth & Stan & 10 & \\
\botrule
\end{tabular}
\caption{Model Parameters: Parameter values and ranges in sensitivity analysis. Type refers to the component of the model where the parameter is used.}\label{param_table}%
\end{table}

We use brms \cite{burkner2018a} and stan \cite{rstan} to fit the model. The parameters used are summarised in Table \ref{param_table}. We used 2000 warmup samples and 2000 iterations per chain, for 4 chains.

MCMC convergence statistics are available in supplementary information \ref{convergence_stats}.

We also provide sensitivity analysis in the supplementary information \ref{sensitivity_analysis}.

\subsection{Identification Strategy} \label{identification}

We are interested in the effects of shocks to response variables in our model. We can compute the impulse response function to gain insights into shocks. The problem described in the previous section is over-identified, with respect to analysing shocks to the system, and we make the following assumptions to enable identification.

\begin{enumerate}
    \item Shocks to $log R$ are contemporaneously unaffected by shocks to $log ED$, $\Delta$ GDP and $\Delta$ Transit. $log R$ is driven by contact rates between individuals who  calibrate their willingness to interact on past observations of excess deaths, travel and economic activity. Contemporaneous information is not yet available to individuals. 
    \item $log ED$ is contemporaneously unaffected by shocks to $\Delta$ GDP and $\Delta$ Transit. $log ED$ is only affected by interactions arising from economic activity or mobility of the {\it previous} time period, as person-to-person contact is required to transmit the disease, and hence neither contemporaneous economic activity nor Transit impact Excess Deaths. 
    \item $\Delta$ GDP is contemporaneously unaffected by shocks to $\Delta$ Transit. Individuals may alter their travel behaviour as a function of the observed economic opportunity / work availability of the previous time period. 
    \item $\Delta$ Transit can be contemporaneously affected by shocks to any of the variables.
\end{enumerate}	

This identification corresponds to the Cholesky decomposition of $\Sigma_u$ which is a zero short run restriction. This identification strategy is implemented / represented by the Orthogonal Impulse Response Function (OIRF) and one of the standard identification strategies for Structural Vector Auto-regressive models. We provide more details in the supplementary information \ref{si_identification}.   

The OIRF has a strict dependence on the ordering of the variables, as set out in the identification strategy. An alternative identification strategy is the Generalised Impulse Response Function (GIRF) \cite{pesaran1998generalized}, which imposes a stricter requirement that assumes that no variable is affected by shocks of the others contemporaneously. This is a very strong assumption and only holds if and only if $\Sigma_u$ is diagonal. In our study $\Sigma_u$ is not diagonal but the diagonal terms dominate. We provide the GIRF for our model as well and observe that it is close to the OIRF (see Figure \ref{oirf_girf_plot}). 

We do not consider NPIs in the computation of the impulse response function, which assumes NPIs are not impacted by shocks. To validate this assumption, we compute the model without NPIs and observe that the estimates and the impulse response function are broadly the same (see section \ref{si_exogeneity} and \ref{si_identification}).

\subsection{Exogeneity of NPIs, vaccinations and dominant variant}\label{exogeneity}

We consider NPIs (of the previous time period) to be exogenous, fixed effects, in the model. This is a significant assumption based on a number of observations. The policy response varies significantly between countries, both in magnitude and timing, see Figure \ref{npi_plot}. The imposition of NPIs may have been more closely linked to the pandemic during the early parts of the pandemic (March-September 2020) but this link has weakened after its initial phase. Changing policy objectives and broad debate on interventions further added variability to have NPIs which had been imposed. 

Similarly we consider dominant variant, which is a function of the evolution of the virus, and vaccination, which is driven by supply of the vaccine, to be exogeneous too. The evolution of the virus, and which variant dominates, is independent of any intervention and the coefficients represent the net effect over that time period with respect to the response variable.  

For validation, we compare the full model and a model that omits all exogeneous variables (see Table \ref{tab:exo_comp_table}) and note that the magnitude of the estimates is approximately the same. Further, we do not see a change of sign from significantly negative to significantly positive. We also consider the OIRFs of both models [Figure \ref{oirf_comparison_plot}] and note that they are very similar leading to the same conclusions on the interaction of the response variables.

\bmhead{Supplementary information}
The data and code are available on \href{https://github.com/cm401/covid_eco_epi_var}{Github}

\bmhead{Acknowledgments}
We acknowledge the Schmidt Foundation for research funding (grant code 6-22-63345), the UK Medical Research Council (MRC) for Centre funding (reference MR/R015600/1), Community Jameel for research funding and the National Institute for Health Research (NIHR) for support for the Health Research Protection Unit for Modelling and Health Economics, a partnership between Public Health England, Imperial College London and LSHTM (grant code NIHR200908). 

\section*{Declarations}

\begin{itemize}
\item Funding
\item Conflict of interest/Competing interests (check journal-specific guidelines for which heading to use)
\item Ethics approval 
\item Consent to participate
\item Consent for publication
\item Availability of data and materials
\item Code availability 
\item Authors' contributions
\end{itemize}

\bibliography{sn-bibliography}

\newpage
\appendix

\begin{center}
    {\LARGE \bf{Supplementary Information}}
    \newline
    \newline
    \bf{Estimating the interaction of transmission intensity, mortality and the economy: a retrospective analysis of the COVID-19 pandemic} 
\end{center}

\renewcommand\thefigure{\thesection.\arabic{figure}}    
\renewcommand\thetable{\thesection.\arabic{table}}    
\setcounter{figure}{0}  
\setcounter{table}{0}  
\setcounter{page}{1}

\newcommand{\beginsupplement}{%
         \setcounter{table}{0}
        \renewcommand{\thetable}{S\arabic{table}}%
        \setcounter{figure}{0}
        \renewcommand{\thefigure}{S\arabic{figure}}%
        \renewcommand \thepart{}
        \renewcommand \partname{}
     }
     
\beginsupplement

\part{} %
\parttoc %
     
\newpage
\section{SARS-CoV-2 Data \& NPI descriptions}

We consider data for the SARS-CoV-2 pandemic for 25 European countries from 1 Jan 2020 to 31 Dec 2021. The start of the data availability varies by country as a function of the start of the pandemic and data collection in that country.  

\subsection{Response Variables, by country, over the full period}

\begin{figure}[!htp]
\centering
\includegraphics[width=1\textwidth]{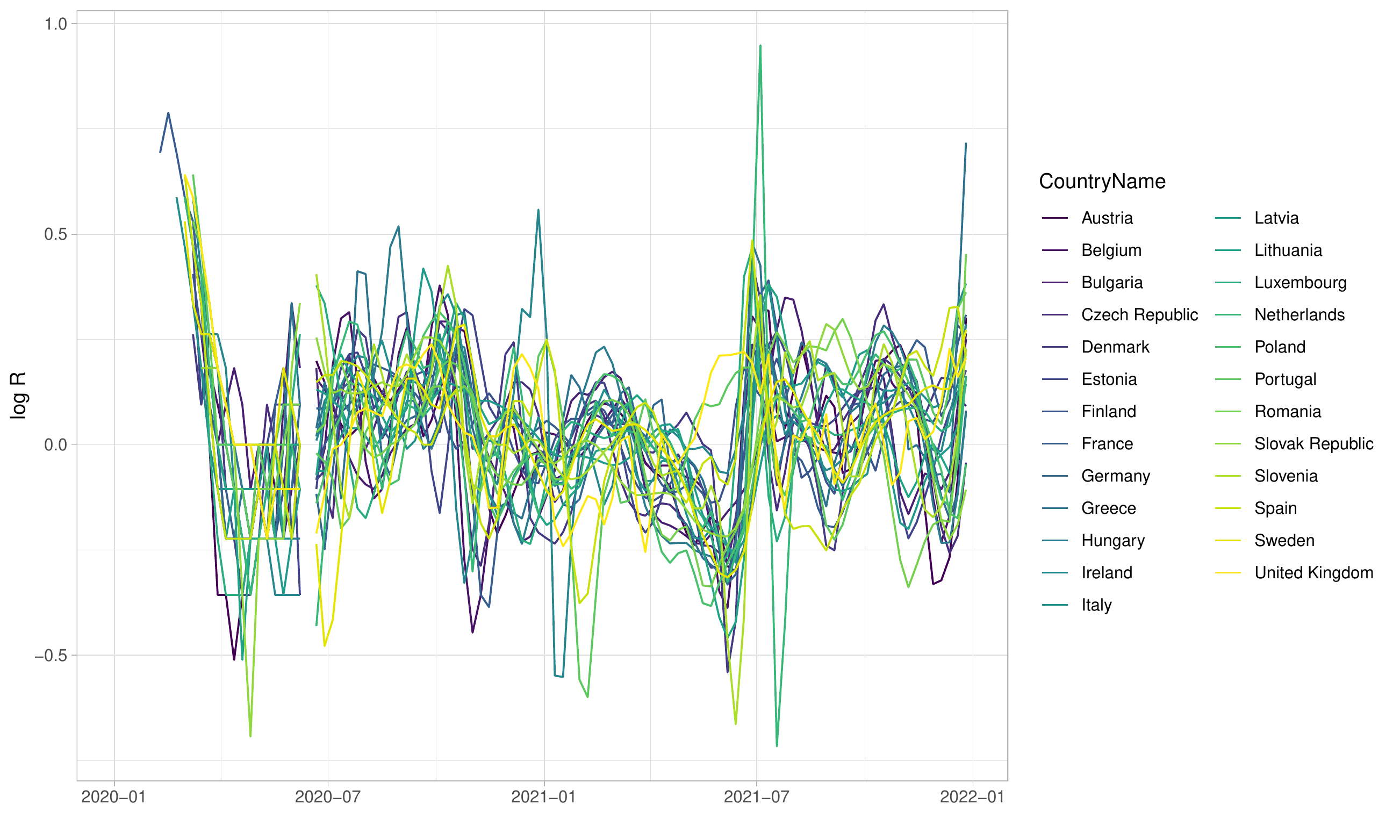}
\caption{log R (transmission intensity) over the course of 2020-21. We can observe heterogeneity across both countries and time.}\label{response_variables_plot_logR}
\end{figure}

\begin{figure}[!htp]
\centering
\includegraphics[width=1\textwidth]{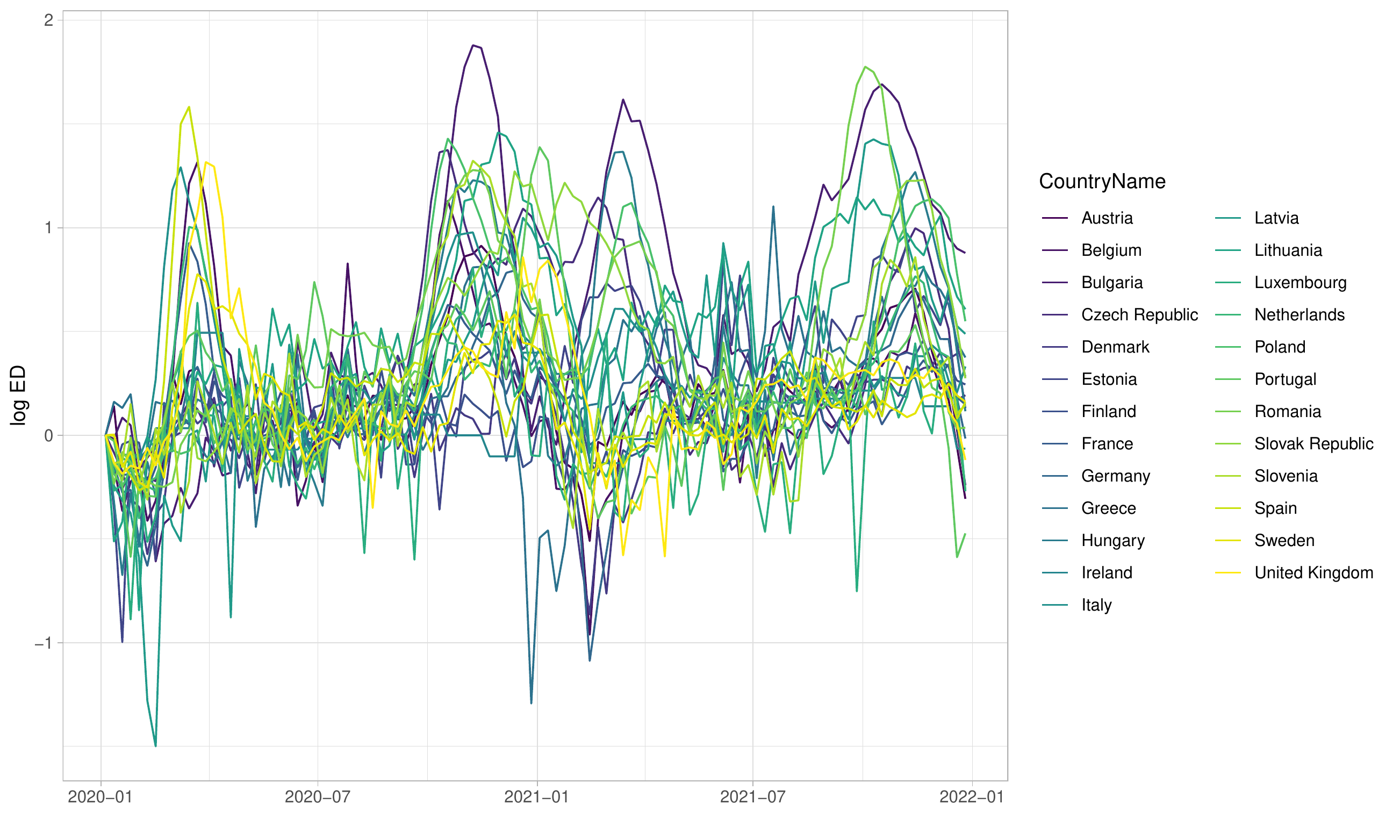}
\caption{log excess deaths over the course of 2020-21. We can observe heterogeneity across both countries and time.}\label{response_variables_plot_logED}
\end{figure}

\begin{figure}[!htp]
\centering
\includegraphics[width=1\textwidth]{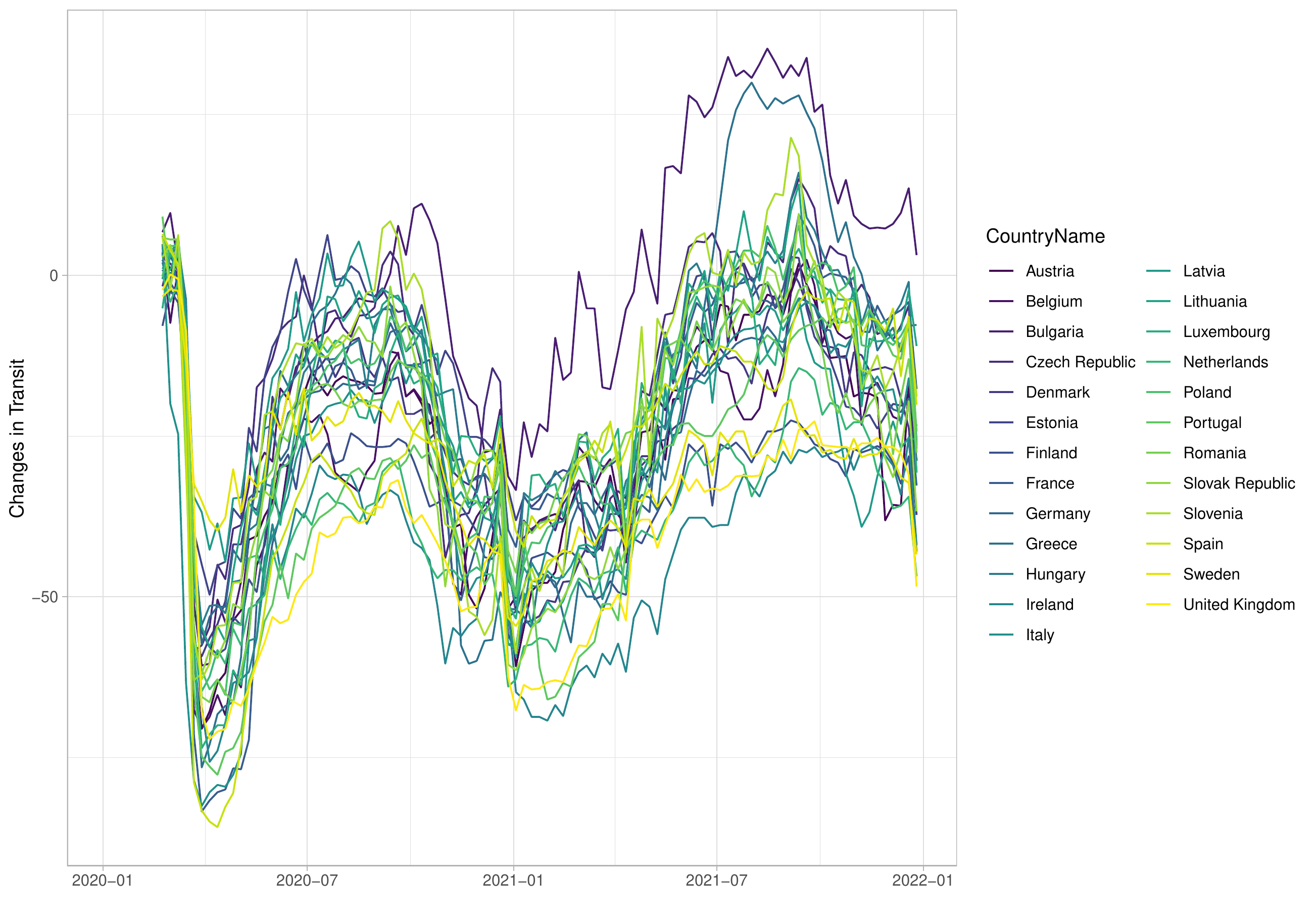}
\caption{Changes in transit (Google mobility) over the course of 2020-21. We can observe heterogeneity across both countries and time.}\label{response_variables_plot_transit}
\end{figure}

\begin{figure}[!htp]
\centering
\includegraphics[width=1\textwidth]{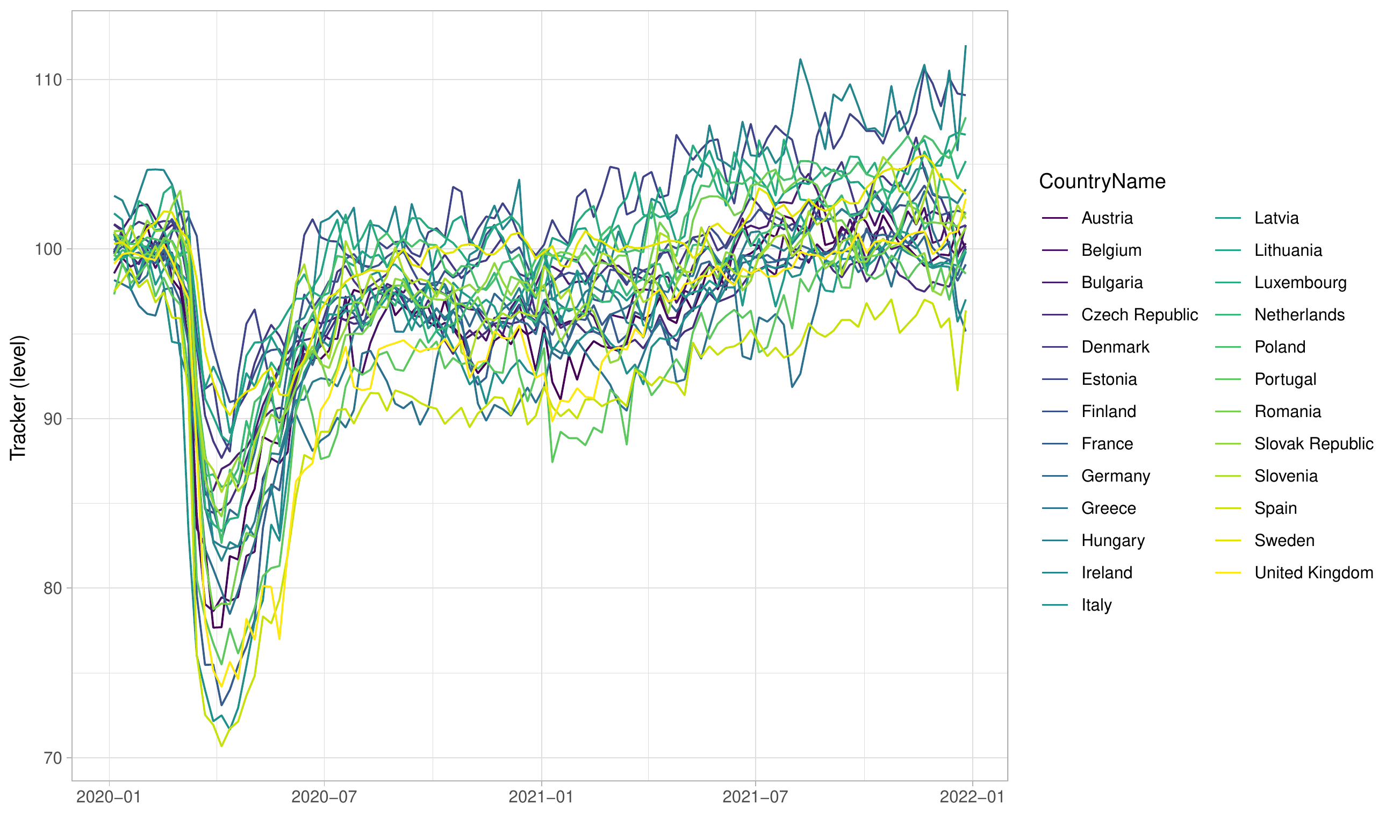}
\caption{Indexed GDP over the course of 2020-21. We can observe heterogeneity across both countries and time.}\label{response_variables_plot_GDP}
\end{figure}

\begin{figure}[!htp]
\centering
\includegraphics[width=0.75\textwidth]{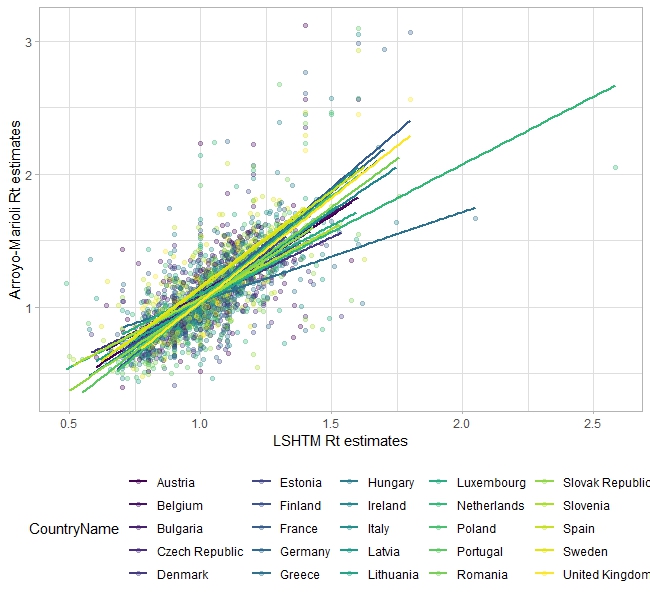}
\caption{$R_t$ comparison of \cite{abbott2020} and \cite{arroyo-marioli2021} }\label{rt_comparison_plot}
\end{figure}

\newpage

\subsection{Oxford Covid Government Response - Measures across Containment \& Health policies, Economic policies and Health System policies}

NPI data is obtained from the Oxford Blavatnik School of Government Covid-19 Government Response Tracker (OxCGRT) \cite{Oxford_BSG}. OxCGRT provides systematic data on interventions implemented by governments on a daily basis, since the start of the pandemic. The policy actions are split into 3 categories: 8 containment and closure policy indicators (C1-C8), 4 economic policy indicators (E1-E4) and 8 health system policy indicators (H1-H8). For each measure a score is available that reflects the severity / scale of the restriction, see Tables \ref{OxCGRT_codebook_table_C}, \ref{OxCGRT_codebook_table_E}, \ref{OxCGRT_codebook_table_H}. 

\singlespacing

\begin{sidewaystable}      %
\sidewaystablefn%
\begin{tabular}{@{}lccl@{}}
\toprule
ID & Name & Description & Coding\\
\midrule
C1 & School closing & Record closings  & 1 - recommend closing or open with alterations\\
   & & of schools and &  2 - require closing (only some levels or categories)\\
   & & universities &  3 - require closing all levels\\
C2 & workplace closure & Record closings of &  1 - recommend closing or all businesses open with changes \\
   & & workplaces &  2 - require closing (only some levels or categories) \\
   & & & 3 - require closing for all-but-essential workplaces \\
C3 & Cancel public events & Record cancelling & 1 - recommend cancelling\\
   & & public events & 2 - require cancelling\\
C4 & Restrictions on gatherings & Record limits on & 1 - restrictions above 1000 people \\
   & & gatherings & 2 - restrictions between 101-1000 people \\
   & & & 3 - restrictions between 11-100 people \\
   & & & 4 - restrictions on 10 people or less\\
C5 & Close public transport & Record closing of &  1 - recommend closing \\
   & & public transport &  2 - require closing\\
C6 & Stay at home requirement & Record orders to "shelter- & 1 - recommend not leaving  house\\
   & & in-place" and otherwise & 2 - require not leaving house with exceptions\\
   & & confine to the home & \\
C7 & Restrictions on & Record restrictions & 1 - recommend not to travel between regions/cities \\
   & internal movement & on internal movement & 2 - internal movement restrictions in place\\
   & & between cities/regions & 3 - require not leaving house with minimal exceptions\\
C8 & International & Record restrictions  & 1 - screening arrivals \\
   & travel & on international & 2 - quarantine arrivals from some or all regions \\
   & controls & travel\footnote{Note: this records policy for foreign travellers, not citizens} & 3 - ban arrivals from some regions \\
   & & & 4 - ban on all regions or total border closure\\
\botrule
\end{tabular}
\caption{OxCGRT Codebook Containment and closure policies \cite{Oxford_BSG}: Missing data will be represented as blank in the database, Coding 0 will be applied if no measure was in place at that moment in time. All Measures are at Ordinal Scale. }\label{OxCGRT_codebook_table_C}%
\end{sidewaystable}

\begin{sidewaystable}     %
\sidewaystablefn%
\begin{tabular}{@{}lcccl@{}}
\toprule
ID & Name & Description & Measure & Coding\\
\midrule
E1 & Income Support & Record if the government & Ordinal & 1 - government is replacing less than 50\% of lost salary \\
   & & is providing direct cash    & scale & (or if a flat sum, it is less than 50\% median salary)\\
   & & payments to people who lose & & 2 - government is replacing 50\% or more of lost salary \\
   & & their jobs or cannot work.\footnote{Note: only includes payments to firms if explicitly linked to payroll/salaries} & & (or if a flat sum, it is greater than 50\% median salary)\\
E2 & Debt/contract & Record if the government is & Ordinal & 1 - narrow relief, specific to one kind of contract \\
   & relief & freezing financial obligations   & scale & 2 - broad debt/contract relief \\
   & & for households & & \\
E3 & Fiscal measures & Announced economic & USD &  Record monetary value in USD of fiscal \\
   & & stimulus spending & & stimuli, includes any spending or tax cuts \\
   & & & & NOT included in E4, H4 or H5 \\
E4 & International & Announced offers of Covid-19  & USD & Record monetary value in USD \\
   & Support & related aid spending  & & \\
   & & to other countries & &  \\
\botrule
\end{tabular}
\caption{OxCGRT Codebook economic policies \cite{Oxford_BSG}: Missing data will be represented as blank in the database, Coding 0 will be applied if no measure or support was in place at that moment in time. }\label{OxCGRT_codebook_table_E}%
\end{sidewaystable}

\begin{sidewaystable}     %
\sidewaystablefn%
\begin{tabular}{@{}lcccl@{}}
\toprule
ID & Name & Description & Measure & Coding\\
\midrule
H1 & Public information & Record presence of  & Ordinal & 1 - public officials urging caution about Covid-19 \\
   & campaigns  & public info campaigns & scale & 2 - coordinated public information campaign\\
H2 & Testing policy & Record government policy & Ordinal & 1 - only those who both (a) have symptoms AND \\
   & & on who has access to & & (b) meet specific criteria\\
   & & testing  & scale & 2 - testing of anyone showing Covid-19 symptoms\\
   & & & & 3 - open public testing \\
H3 & Contact tracing & Record government policy  & Ordinal & 1 - limited contact tracing; not done for all cases \\
   & & on contact tracing after & scale & 2 - comprehensive contact tracing; \\
   & & a positive diagnosis\footnote{Note: we are looking for policies that would identify all people potentially exposed to Covid-19; voluntary bluetooth apps are unlikely to achieve this} & & done for all identified cases \\
H4 & Emergency investment & Announced spending & USD & Record monetary value in USD\\ 
   & in healthcare  & on healthcare system \footnote{Note: only record amount additional to previously announced spending} & & \\
H5 & Investment in vaccines & Announced public spending on & USD & Record monetary value in USD\\
   & & Covid-19 vaccine development & & \\
H6 & Facial coverings & Record policies on & Ordinal &  1 - Recommended\\
   & & the use of facial coverings   & scale & 2 - Required in some specified shared/public spaces\\
   & & outside the home & & 3 - Required in all shared/public spaces\\
   & & & & 4 - Required outside the home at all times\\
H7 & Vaccination policy & Record policies & Ordinal & 1 - Availability for ONE group\footnote{Groups: key workers/ clinically vulnerable groups (non elderly) / elderly groups} \\
   & & for vaccine delivery & scale & 2 - Availability for TWO group\\
   & & for different groups & & 3 - Availability for ALL group \\
   & & & & 4 - Available to ALL groups + some others\\
   & & & & 5 - Universally available\\
H8 & Protection of & Record policies for protecting  & Ordinal & 1 - Recommended restriction measures in LTCFs\\ 
   & elderly people & elderly people in LT Care & scale & 2 - Narrow restrictions measures in LTCFs\\ 
   & & Facilities/community setting & & 3 - Extensive restrictions measures in LTCFs\\
\botrule
\end{tabular}
\caption{OxCGRT Codebook Health system policies \cite{Oxford_BSG}: Missing data will be represented as blank in the database, Coding 0 will be applied if no measure was in place at that moment in time.}\label{OxCGRT_codebook_table_H}%
\end{sidewaystable}

\begin{figure}[!htp]
\centering
\includegraphics[width=1\textwidth]{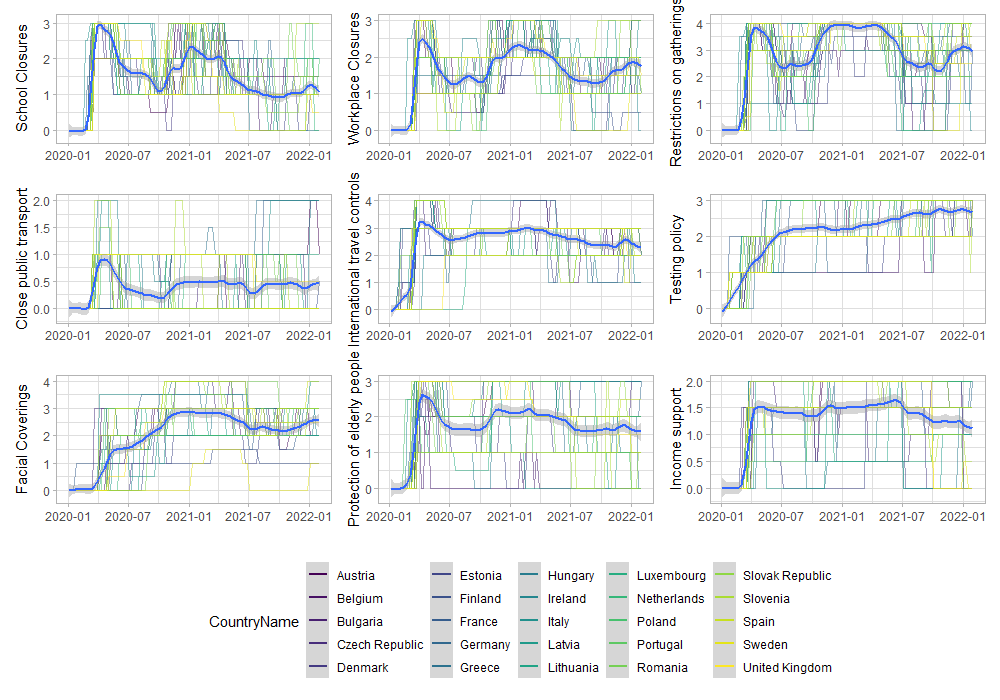}
\caption{Non-pharmaceutical Interventions: Time-series of NPIs over the cause of the pandemic by country. The blue line is the smoothed pan-European average for each NPI.}\label{npi_plot}
\end{figure}

\begin{figure}[!htp]
\centering
\includegraphics[width=1\textwidth]{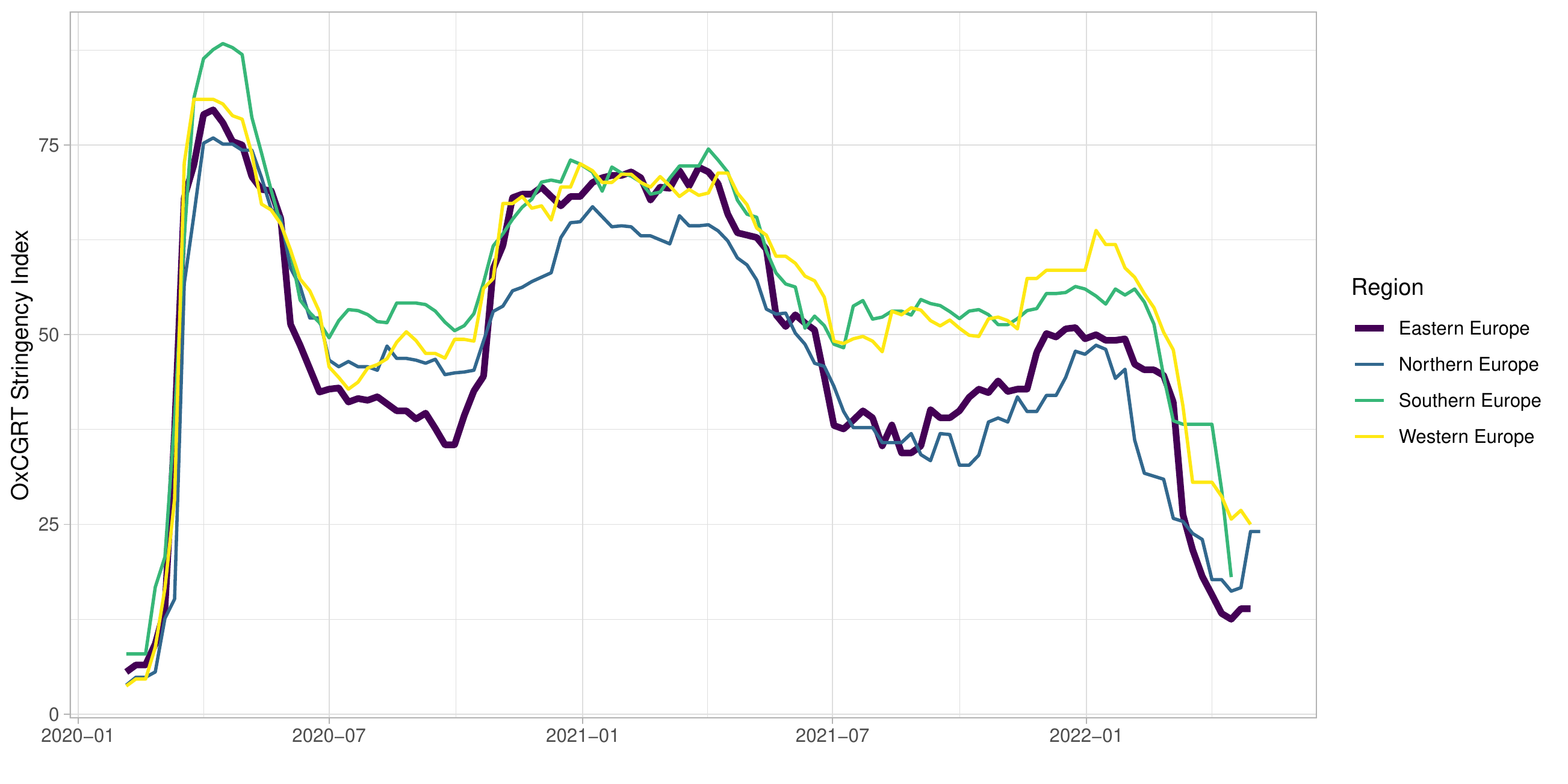}
\caption{Oxford Covid Government Responds Tracker NPI Stringency by European sub-region.}\label{npi_eastern_europe_plot}
\end{figure}

\newpage

\subsection{Vaccination \& Variant data}

 Our World In Data \cite{mathieu2021global} provides a range of vaccination statistics. We use the total number of vaccine doses administered over time. dividing by the total population \cite{eurostat_data} to calculate the number of doses delivered per capita.

\begin{figure}[!htp]
\centering
\includegraphics[width=1\textwidth]{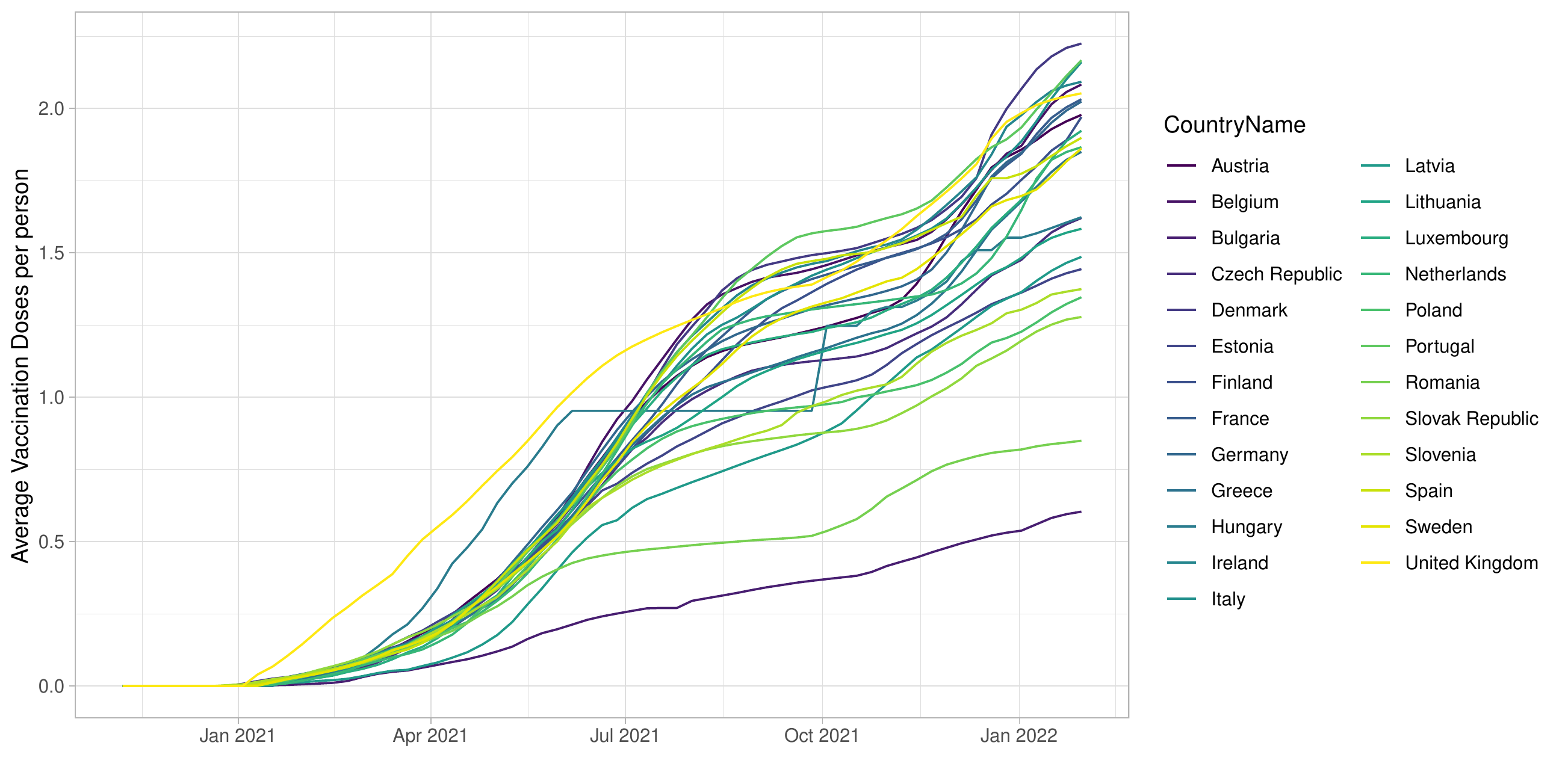}
\caption{Vaccinations: Average vaccination per person (total number of vaccines delivered divided by total population).}\label{vaccination_plot}
\end{figure}

\begin{figure}[!htp]
\centering
\includegraphics[width=1\textwidth]{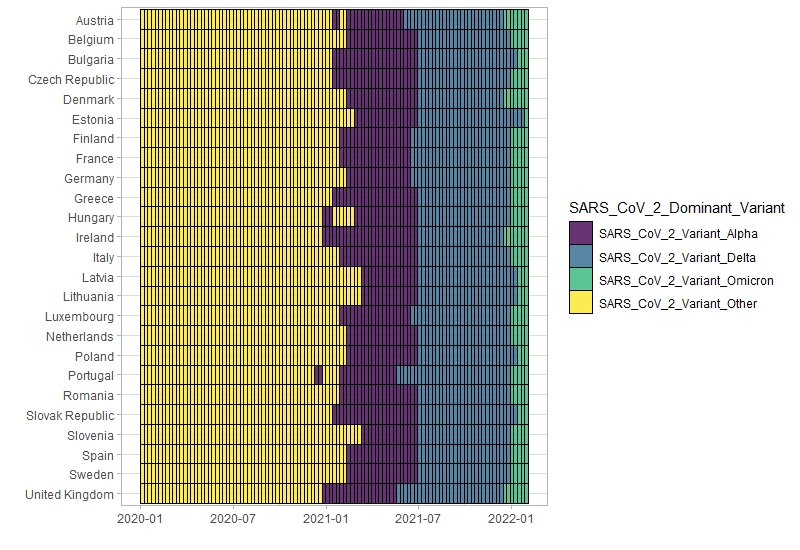}
\caption{SARS-CoV-2 Dominant Variant: Largest Variant as measured by sequenced samples over time in each country.}\label{dominant_variant}
\end{figure}

SARS-CoV-2 variant data is sourced from \cite{covariant}, which provides a summary view by country and week based on GISAID data. We utilise the Nextstrain Clade, Pango Lineage and WHO Label mapping to map all Nextstrain Clades into WHO Labels, for use in the model. For each week we pick the dominant strain in each country as the strain with the majority of sequenced samples (Figure \ref{dominant_variant}). We observe a data gap for Hungary, which is missing weeks 9-45 in 2021. We estimate data for Hungary during this time period by using the average of the data from surrounding countries, weighted by border length.

\clearpage
\subsection{GDP Impact}

\begin{figure}[!htp]
\centering
\includegraphics[width=1\textwidth]{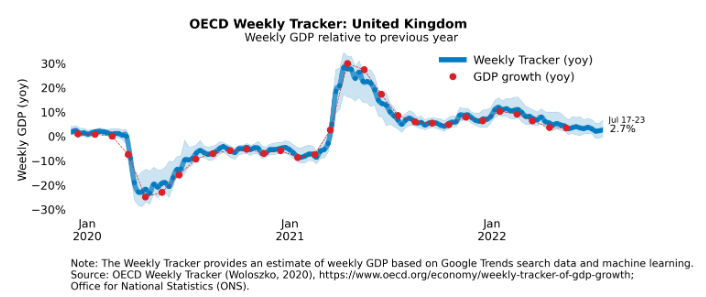}
\caption{OECD GDP Nowcast validation: Comparsion of GDP Nowcast (YoY) and  GDP (YoY) observations from the ONS for the United Kingdom.   [Source: \href{https://www.oecd.org/economy/weekly-tracker-of-gdp-growth/}{OECD Link}}]\label{oecd_nowcast_validation}
\end{figure}

\begin{figure}[!htp]
\centering
\includegraphics[width=1\textwidth]{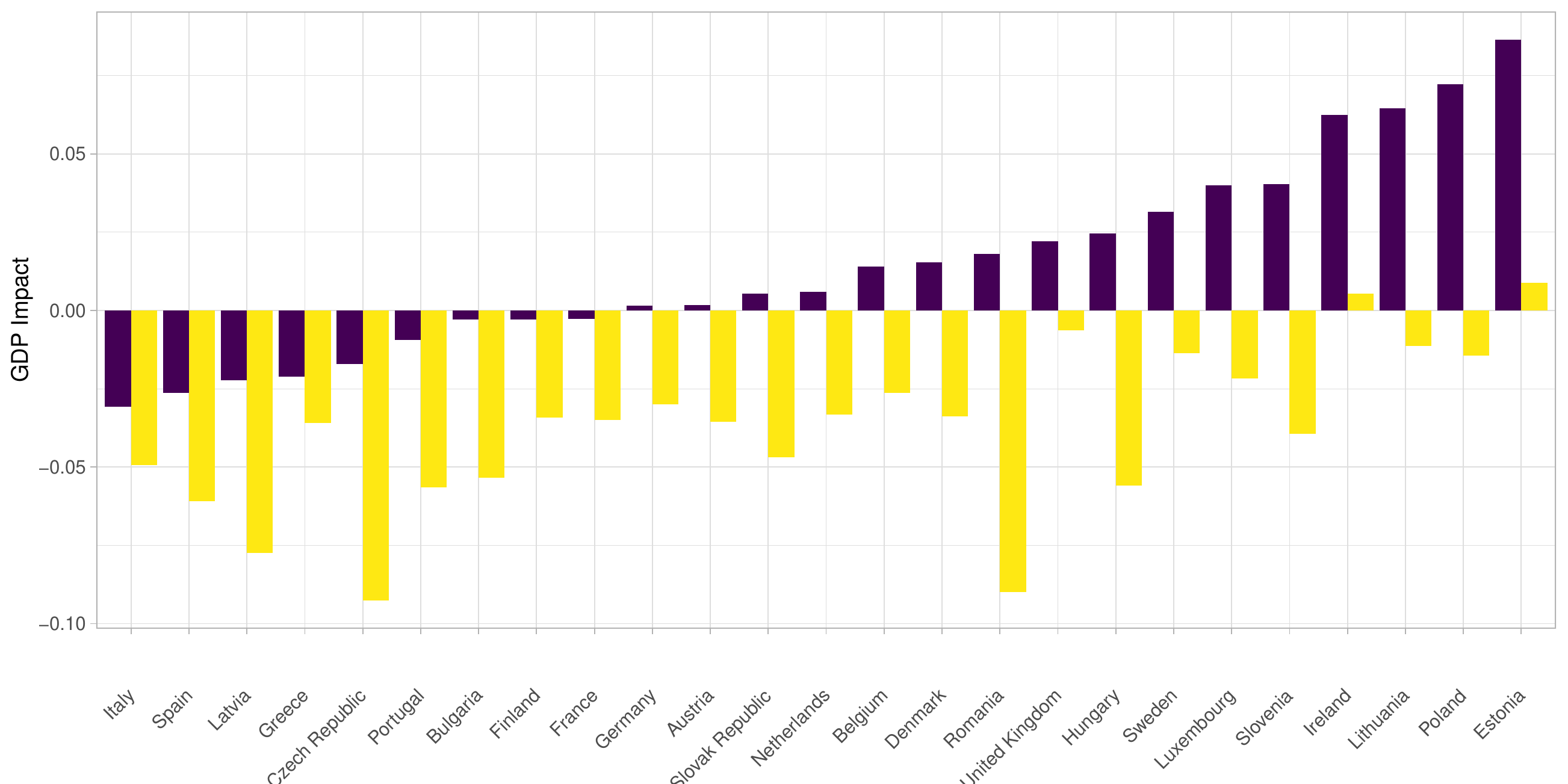}
\caption{GDP Impact: We consider the overall GDP growth over the full 2 year period of 2020-21. The purple bars are the observed GDP and the yellow bars are the GDP {\it including} the counterfactual.}\label{gdp_impact}
\end{figure}

\clearpage

\subsection{School Closure data \& impact}

\begin{figure}[!htp]
\centering
\includegraphics[width=1\textwidth]{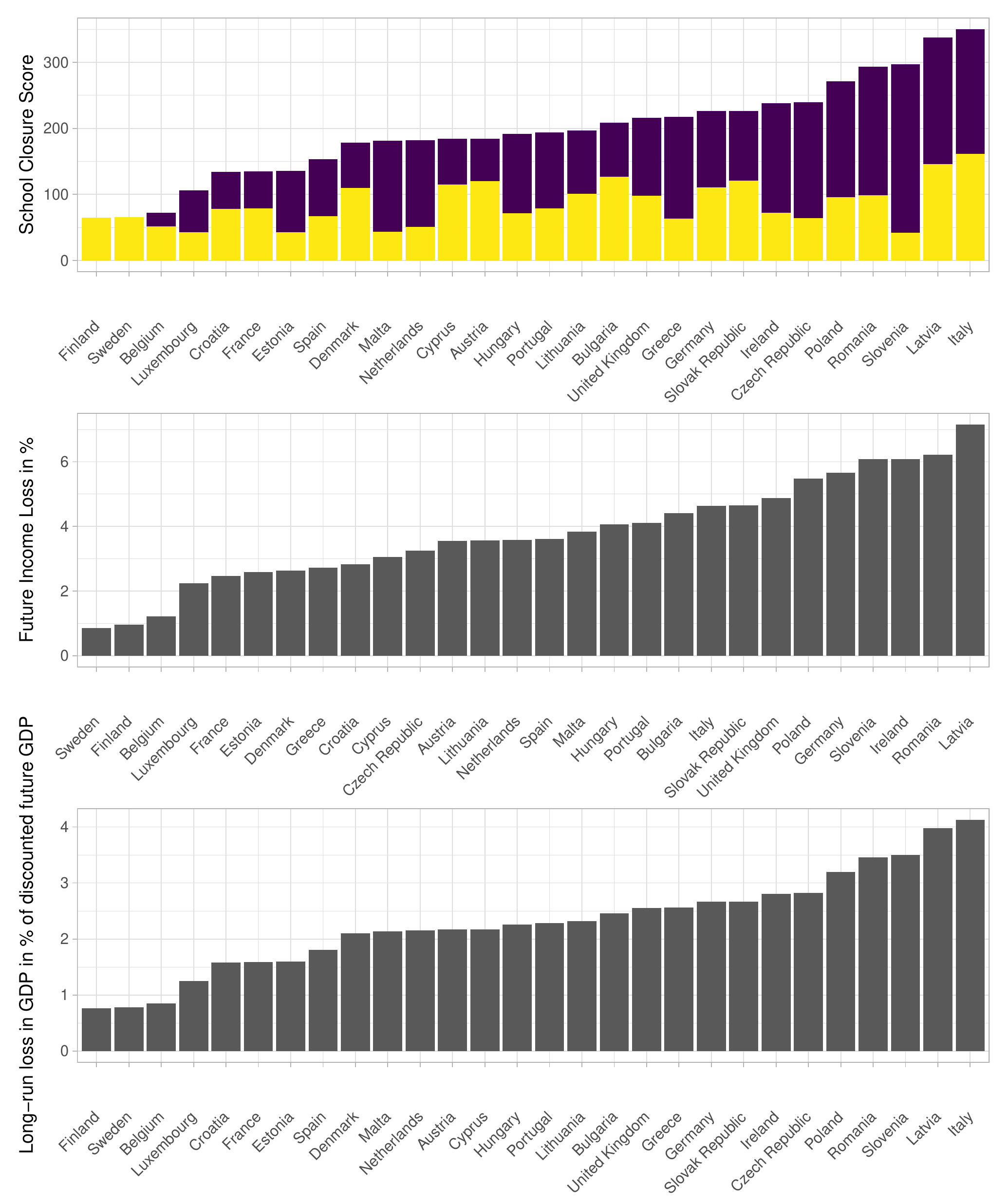}
\caption{School Closures: The top panel reports the School Closure score (purple component is equal to the number of days of full school closures, yellow component is a proxy for part shut downs). The middle panel reports the estimated future income loss for impacted students and the bottom panel reports the Long-run loss in GDP as a percentage of discounted future GDP.\\
{\it Source:} OECD \cite{school_closure_oecd}}\label{school_closures_plot}
\end{figure}

\newpage
\section{Model outputs \& analytics}

In this section we report detailed outputs from the model presented in this paper.

\subsection{Coefficient Estimates}

We provide the coefficient estimates for the model in 5 tables:

\begin{enumerate}
    \item Estimates of the VAR coefficients $\Phi_{i,j,k=1}$ in Table \ref{tab:VAR_coef}
    \item Estimates of the coefficients for NPI levels $\lambda$ in Table \ref{tab:NPI_level_coef}
    \item Estimates of the coefficients for NPI changes $\delta$ in Table \ref{tab:NPI_changes_coef}
    \item Estimates for the Dominant Varient coefficients $\nu$ in Table \ref{tab:Dominant_variant_coef}
    \item Estimates for Vaccination coefficients $\psi$ in Table \ref{tab:vaccination_coef}
\end{enumerate}

\singlespacing

\begin{table}[!htp]
 \centering
\begin{tabular}[t]{lllll}
\toprule
Variable & Estimate & Est.Error & CrI Lower & CrI Upper\\
\midrule
log R, log R l1 & 0.757 & 0.017 & 0.725 & 0.790\\
log R, log ED l1 & -0.040 & 0.007 & -0.054 & -0.026\\
log R, $\Delta$ GDP l1 & 0.003 & 0.017 & -0.032 & 0.036\\
log R, $\Delta$ Transit l1 & 0.103 & 0.043 & 0.019 & 0.187\\
\addlinespace
log ED, log R l1 & 0.271 & 0.030 & 0.213 & 0.329\\
log ED, log ED l1 & 0.856 & 0.014 & 0.828 & 0.884\\
log ED, $\Delta$ GDP l1 & -0.008 & 0.031 & -0.067 & 0.053\\
log ED, $\Delta$ Transit l1 & 0.014 & 0.076 & -0.135 & 0.163\\
\addlinespace
$\Delta$ GDP, log R l1 & -0.241 & 0.027 & -0.295 & -0.189\\
$\Delta$ GDP, log ED l1 & -0.054 & 0.012 & -0.078 & -0.030\\
$\Delta$ GDP, $\Delta$ GDP l1 & 0.046 & 0.028 & -0.008 & 0.100\\
$\Delta$ GDP, $\Delta$ Transit l1 & 0.067 & 0.069 & -0.067 & 0.203\\
\addlinespace
$\Delta$ Transit, log R l1 & -0.055 & 0.010 & -0.074 & -0.036\\
$\Delta$ Transit, log ED l1 & -0.025 & 0.004 & -0.033 & -0.018\\
$\Delta$ Transit, $\Delta$ GDP l1 & 0.135 & 0.010 & 0.115 & 0.156\\
$\Delta$ Transit, $\Delta$ Transit l1 & -0.113 & 0.025 & -0.163 & -0.063\\
\bottomrule
\end{tabular}
\caption{Estimates for VAR coefficients $\Phi_{i,j,k=1}$ of the model}
    \label{tab:VAR_coef}
\end{table}

\begin{table}[]
    \centering
    \begin{tabular}[t]{lllll}
\toprule
Variable & Estimate & Est.Error & CrI Lower & CrI Upper\\
\midrule
log R lvl Schools Closing & -0.006 & 0.004 & -0.014 & 0.002\\
log R lvl workplace closure & -0.012 & 0.005 & -0.021 & -0.003\\
log R lvl Restrictions on gatherings & -0.006 & 0.003 & -0.012 & -0.001\\
log R lvl Close public transport & 0.007 & 0.005 & -0.003 & 0.017\\
log R lvl International travel controls & 0.003 & 0.004 & -0.004 & 0.011\\
log R lvl Testing policy & 0.007 & 0.005 & -0.003 & 0.016\\
log R lvl Facial Coverings & 0.010 & 0.003 & 0.004 & 0.015\\
log R lvl Protection of elderly people & -0.007 & 0.003 & -0.013 & -0.000\\
log R lvl Income support & 0.002 & 0.004 & -0.006 & 0.010\\
\addlinespace
log ED lvl Schools Closing & -0.010 & 0.008 & -0.026 & 0.006\\
log ED lvl workplace closure & -0.002 & 0.009 & -0.019 & 0.015\\
log ED lvl Restrictions on gatherings & 0.002 & 0.005 & -0.008 & 0.012\\
log ED lvl Close public transport & 0.005 & 0.010 & -0.015 & 0.024\\
log ED lvl International travel controls & -0.014 & 0.007 & -0.028 & -0.001\\
log ED lvl Testing policy & -0.005 & 0.009 & -0.023 & 0.013\\
log ED lvl Facial Coverings & 0.011 & 0.005 & 0.000 & 0.021\\
log ED lvl Protection of elderly people & -0.001 & 0.006 & -0.014 & 0.011\\
log ED lvl Income support & -0.007 & 0.008 & -0.023 & 0.009\\
\addlinespace
$\Delta$ GDP lvl Schools Closing & 0.007 & 0.007 & -0.007 & 0.020\\
$\Delta$ GDP lvl workplace closure & 0.002 & 0.008 & -0.013 & 0.018\\
$\Delta$ GDP lvl Restrictions on gatherings & 0.004 & 0.005 & -0.005 & 0.013\\
$\Delta$ GDP lvl Close public transport & -0.001 & 0.009 & -0.018 & 0.016\\
$\Delta$ GDP lvl International travel controls & 0.014 & 0.006 & 0.002 & 0.025\\
$\Delta$ GDP lvl Testing policy & 0.007 & 0.008 & -0.008 & 0.024\\
$\Delta$ GDP lvl Facial Coverings & 0.016 & 0.005 & 0.007 & 0.026\\
$\Delta$ GDP lvl Protection of elderly people & 0.015 & 0.006 & 0.005 & 0.027\\
$\Delta$ GDP lvl Income support & 0.023 & 0.008 & 0.009 & 0.039\\
\addlinespace
$\Delta$ Transit lvl Schools Closing & 0.001 & 0.002 & -0.003 & 0.006\\
$\Delta$ Transit lvl workplace closure & -0.000 & 0.003 & -0.006 & 0.005\\
$\Delta$ Transit lvl Restrictions on gatherings & 0.002 & 0.002 & -0.001 & 0.005\\
$\Delta$ Transit lvl Close public transport & 0.001 & 0.003 & -0.005 & 0.007\\
$\Delta$ Transit lvl International travel controls & 0.003 & 0.002 & -0.002 & 0.007\\
$\Delta$ Transit lvl Testing policy & -0.002 & 0.003 & -0.008 & 0.003\\
$\Delta$ Transit lvl Facial Coverings & 0.001 & 0.002 & -0.002 & 0.005\\
$\Delta$ Transit lvl Protection of elderly people & 0.002 & 0.002 & -0.002 & 0.006\\
$\Delta$ Transit lvl Income support & 0.002 & 0.003 & -0.003 & 0.006\\
\bottomrule
\end{tabular}
\caption{Estimates for NPI Level ($\lambda$) coefficients}
    \label{tab:NPI_level_coef}
\end{table}

\begin{table}[]
    \centering
    \begin{tabular}[t]{lllll}
\toprule
Variable & Estimate & Est.Error & CrI Lower & CrI Upper\\
\midrule
log R $\Delta$ Schools Closing & -0.003 & 0.007 & -0.017 & 0.011\\
log R $\Delta$ workplace closure & -0.011 & 0.008 & -0.026 & 0.005\\
log R $\Delta$ Restrictions on gatherings & -0.004 & 0.005 & -0.014 & 0.006\\
log R $\Delta$ Close public transport & -0.013 & 0.014 & -0.040 & 0.015\\
log R $\Delta$ International travel controls & 0.007 & 0.008 & -0.008 & 0.022\\
log R $\Delta$ Testing policy & -0.003 & 0.011 & -0.024 & 0.018\\
log R $\Delta$ Facial Coverings & -0.004 & 0.009 & -0.022 & 0.013\\
log R $\Delta$ Protection of elderly people & -0.000 & 0.007 & -0.015 & 0.013\\
log R $\Delta$ Income support & 0.014 & 0.011 & -0.008 & 0.037\\
\addlinespace
log ED $\Delta$ Schools Closing & -0.007 & 0.013 & -0.032 & 0.019\\
log ED $\Delta$ workplace closure & 0.005 & 0.014 & -0.023 & 0.033\\
log ED $\Delta$ Restrictions on gatherings & 0.021 & 0.009 & 0.003 & 0.039\\
log ED $\Delta$ Close public transport & -0.001 & 0.025 & -0.052 & 0.049\\
log ED $\Delta$ International travel controls & -0.026 & 0.014 & -0.053 & 0.001\\
log ED $\Delta$ Testing policy & 0.005 & 0.020 & -0.033 & 0.043\\
log ED $\Delta$ Facial Coverings & -0.003 & 0.016 & -0.034 & 0.028\\
log ED $\Delta$ Protection of elderly people & 0.020 & 0.013 & -0.005 & 0.045\\
log ED $\Delta$ Income support & 0.003 & 0.021 & -0.038 & 0.044\\
\addlinespace
$\Delta$ GDP $\Delta$ Schools Closing & -0.026 & 0.012 & -0.049 & -0.003\\
$\Delta$ GDP $\Delta$ workplace closure & -0.002 & 0.013 & -0.027 & 0.022\\
$\Delta$ GDP $\Delta$ Restrictions on gatherings & 0.001 & 0.008 & -0.015 & 0.016\\
$\Delta$ GDP $\Delta$ Close public transport & -0.017 & 0.023 & -0.061 & 0.028\\
$\Delta$ GDP $\Delta$ International travel controls & -0.023 & 0.012 & -0.047 & 0.001\\
$\Delta$ GDP $\Delta$ Testing policy & -0.022 & 0.017 & -0.056 & 0.012\\
$\Delta$ GDP $\Delta$ Facial Coverings & 0.005 & 0.014 & -0.023 & 0.032\\
$\Delta$ GDP $\Delta$ Protection of elderly people & -0.049 & 0.011 & -0.071 & -0.027\\
$\Delta$ GDP $\Delta$ Income support & -0.007 & 0.018 & -0.043 & 0.028\\
\addlinespace
$\Delta$ Transit $\Delta$ Schools Closing & -0.019 & 0.004 & -0.028 & -0.011\\
$\Delta$ Transit $\Delta$ workplace closure & -0.026 & 0.005 & -0.036 & -0.017\\
$\Delta$ Transit $\Delta$ Restrictions on gatherings & -0.004 & 0.003 & -0.009 & 0.002\\
$\Delta$ Transit $\Delta$ Close public transport & -0.017 & 0.008 & -0.033 & -0.000\\
$\Delta$ Transit $\Delta$ International travel controls & -0.030 & 0.005 & -0.038 & -0.021\\
$\Delta$ Transit $\Delta$ Testing policy & 0.003 & 0.006 & -0.013 & 0.008\\
$\Delta$ Transit $\Delta$ Facial Coverings & -.003 & 0.005 & -0.013 & 0.008\\
$\Delta$ Transit $\Delta$ Protection of elderly people & -0.007 & 0.004 & -0.015 & 0.001\\
$\Delta$ Transit $\Delta$ Income support & -0.004 & 0.007 & -0.018 & 0.009\\
\bottomrule
\end{tabular}
\caption{Estimates for NPI Changes ($\delta$) coefficients}
    \label{tab:NPI_changes_coef}
\end{table}

\begin{table}[!htp]
    \centering
    \begin{tabular}[t]{lllll}
\toprule
Variable & Estimate & Est.Error & CrI Lower & CrI Upper\\
\midrule
log R WT variant & 0.032 & 0.018 & -0.004 & 0.068\\
log R Alpha variant & -0.027 & 0.008 & -0.042 & -0.012\\
log R Delta variant & 0.033 & 0.013 & 0.007 & 0.059\\
log R Omicron variant & 0.090 & 0.034 & 0.022 & 0.156\\
\addlinespace
log ED WT variant & 0.090 & 0.034 & 0.022 & 0.156\\
log ED Alpha variant & 0.018 & 0.014 & -0.010 & 0.045\\
log ED Delta variant & 0.024 & 0.024 & -0.024 & 0.071\\
log ED Omicron variant & -0.129 & 0.067 & -0.259 & 0.003\\
\addlinespace
$\Delta$ GDP WT variant & -0.143 & 0.035 & -0.214 & -0.077\\
$\Delta$ GDP Alpha variant & -0.040 & 0.013 & -0.065 & -0.015\\
$\Delta$ GDP Delta variant & 0.093 & 0.021 & 0.050 & 0.135\\
$\Delta$ GDP Omicron variant & 0.115 & 0.059 & 0.002 & 0.230\\
\addlinespace
$\Delta$ Transit WT variant & -0.013 & 0.011 & -0.034 & 0.008\\
$\Delta$ Transit Alpha variant & 0.013 & 0.005 & 0.004 & 0.022\\
$\Delta$ Transit Delta variant &  0.015 & 0.008 & -0.001 & 0.030\\
$\Delta$ Transit Omicron variant & -0.050 & 0.021 & -0.092 & -0.009\\
\bottomrule
\end{tabular}
    \caption{Estimates for Dominant Variant coefficients ($\nu$) coefficients}
    \label{tab:Dominant_variant_coef}
\end{table}

\begin{table}[!htp]
    \centering
    \begin{tabular}[t]{lllll}
\toprule
Variable & Estimate & Est.Error & CrI Lower & CrI Upper\\
\midrule
log R vaccination & -0.004 & 0.012 & -0.027 & 0.020\\
log ED vaccination & -0.047 & 0.023 & -0.092 & -0.002\\
$\Delta$ GDP vaccination & -0.028 & 0.020 & -0.068 & 0.012\\
$\Delta$ Transit vaccination & -0.020 & 0.007 & -0.033 & -0.005\\
\bottomrule
\end{tabular}
    \caption{Estimates for Vaccination, ($\psi$) coefficients}
    \label{tab:vaccination_coef}
\end{table}

\subsection{Identification strategy} \label{si_identification}

We assume that all exogeneous factors are not impacted by shocks, which reduces the model to a simple VAR(1) of the form $y_{t,c}= \Phi y_{t-1,c} + u_{t,c}$.

We are interested in the impact of (unit) shocks $\epsilon_{t,c}$ on $y_{t,c}$, which we can represent in matrix form as $u_{t,c}=L \epsilon_{t,c}$, where $L$ is an unrestricted $N\times N$ matrix (recall that N is the length of $y_{t,c}$).

Since $\Sigma_u=\mathbb{E} \left[ u^{\top}u \right]$, we obtain  $\Sigma_u = LL^{\top}$. From this the identification problem is clear, since $\Sigma_u$ is a symmetric matrix.

We have set out an identification strategy in the methods section. The corresponding restriction to $L$ would be a lower triangular matrix. The convenient choice here is to pick $L$ to be the Cholesky decomposition of $\Sigma_u$.

\begin{figure}[!htp]
\centering
\includegraphics[width=1\textwidth]{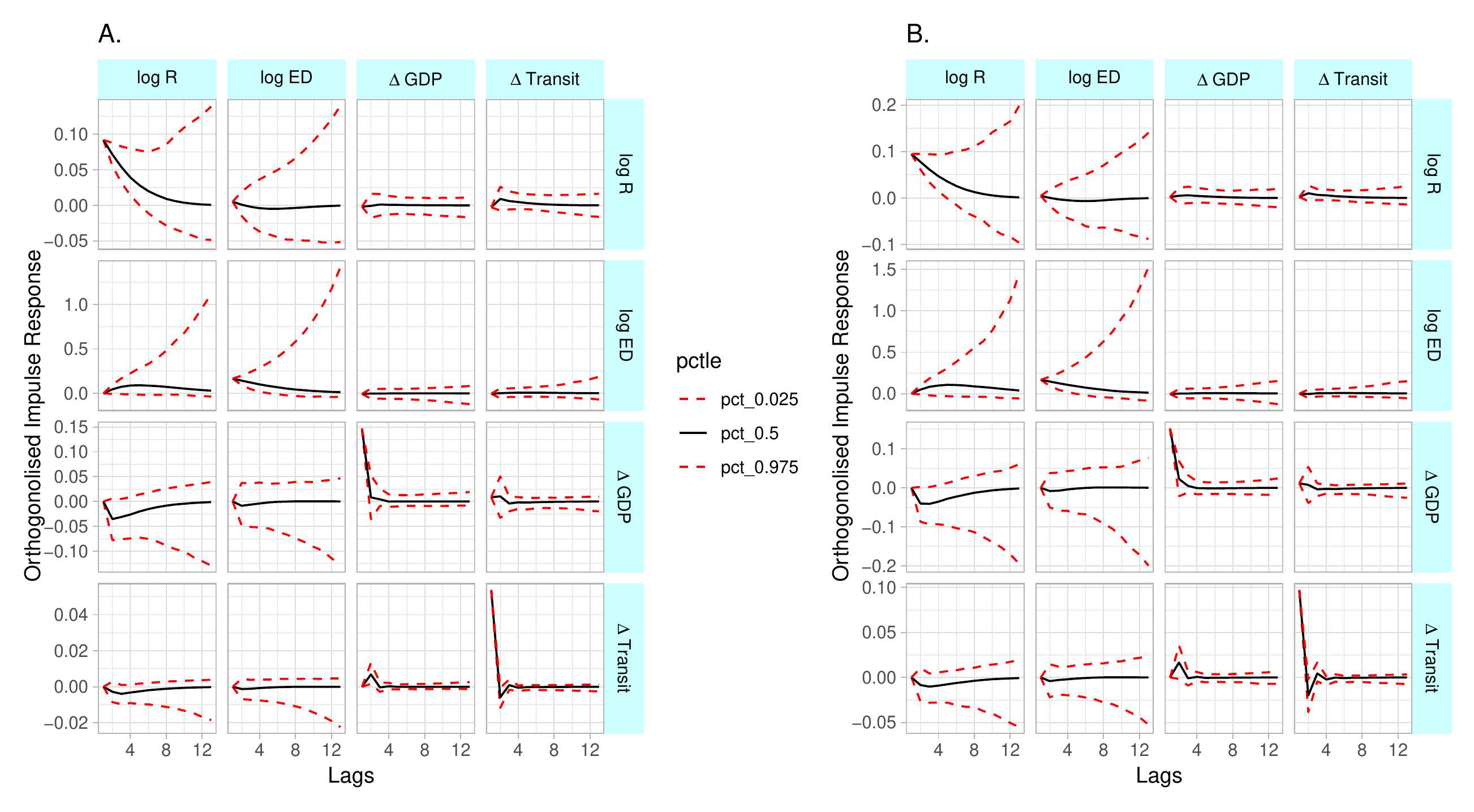}
\caption{Comparison of Orthogonalised Impulse Response functions for full model (A.) and without exogeneous variables (B.)}\label{oirf_comparison_plot}
\end{figure}

\begin{figure}[!htp]
\centering
\includegraphics[width=1\textwidth]{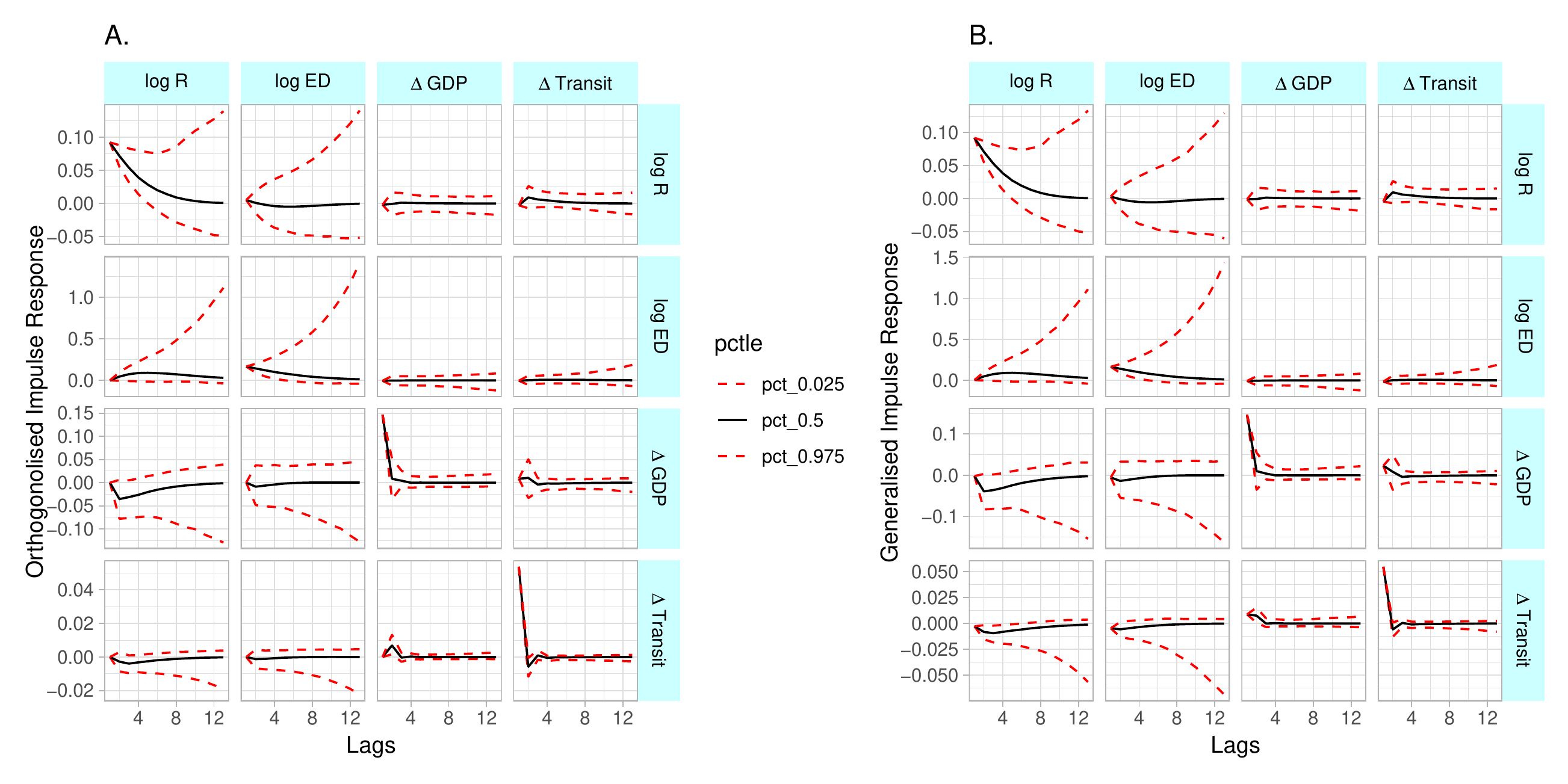}
\caption{Comparison of Orthogonalised Impulse Response functions (A.) and Generalised Impulse Response function (B.)}\label{oirf_girf_plot}
\end{figure}

\subsection{Exogeneity Analysis} \label{si_exogeneity}

\begin{table}[!htp]
\centering
\begin{adjustbox}{width=1\textwidth}
           \begin{tabular}{lllllll}
\toprule
Variable & \multicolumn{3}{c}{{\it With exogeneous factors}} & \multicolumn{3}{c}{{\it No exogenenous factors}} \\
 & Estimate & CrI Lower & CrI Upper & Estimate & CrI Lower & CrI Upper \\
\midrule
log R, log R l1 & 0.757 &  0.725 & 0.790 & 0.811 & 0.783 & 0.840\\
log R, log ED l1 & -0.040  & -0.054 & -0.026 & -0.035 & -0.048 & -0.022\\
log R, $\Delta$ GDP l1 & 0.003 &  -0.032 & 0.036 & 0.038 & 0.007 & 0.068\\
log R, $\Delta$ Transit l1 & 0.103 &  0.019 & 0.187 & 0.108 & 0.061 & 0.155\\
\addlinespace
log ED, log R l1 & 0.271 &  0.213 & 0.329 & 0.284 & 0.233 & 0.335\\
log ED, log ED l1 & 0.856 & 0.828 & 0.884 & 0.869 & 0.843 & 0.894\\
log ED, $\Delta$ GDP l1 & -0.008 & -0.067 & 0.053 & 0.011 & -0.045 & 0.066\\
log ED, $\Delta$ Transit l1 & 0.014 &  -0.135 & 0.163 & -0.014 & -0.099 & 0.069\\
\addlinespace
$\Delta$ GDP, log R l1 & -0.241 & -0.295 & -0.189 & -0.262 &  -0.309 & -0.216\\
$\Delta$ GDP, log ED l1 & -0.054 & -0.078 & -0.030 & -0.043 & -0.063 & -0.023\\
$\Delta$ GDP, $\Delta$ GDP l1 & 0.046 &  -0.008 & 0.100 & 0.139 & 0.090 & 0.189\\
$\Delta$ GDP, $\Delta$ Transit l1 & 0.067 &  -0.067 & 0.203 & 0.063 & -0.014 & 0.137\\
\addlinespace
$\Delta$ Transit, log R l1 & -0.055 &  -0.074 & -0.036 & -0.082 & -0.112 & -0.053\\
$\Delta$ Transit, log ED l1 & -0.025 &  -0.033 & -0.018 & -0.043 &  -0.055 & -0.030\\
$\Delta$ Transit, $\Delta$ GDP l1 & 0.135 & 0.115 & 0.156 & 0.173 &  0.141 & 0.205\\
$\Delta$ Transit, $\Delta$ Transit l1 & -0.113  & -0.163 & -0.063 & -0.210 & -0.259 & -0.162\\
\bottomrule
        \end{tabular}
        \end{adjustbox}
 \caption{Comparison of estimates for VAR coefficients $\Phi_{i,j,k=1}$ of the model with NPIs and without NPIs}
    \label{tab:exo_comp_table}
\end{table}

\subsection{Country specific intercepts}

\begin{figure}[!htp]
\centering
\includegraphics[width=1\textwidth]{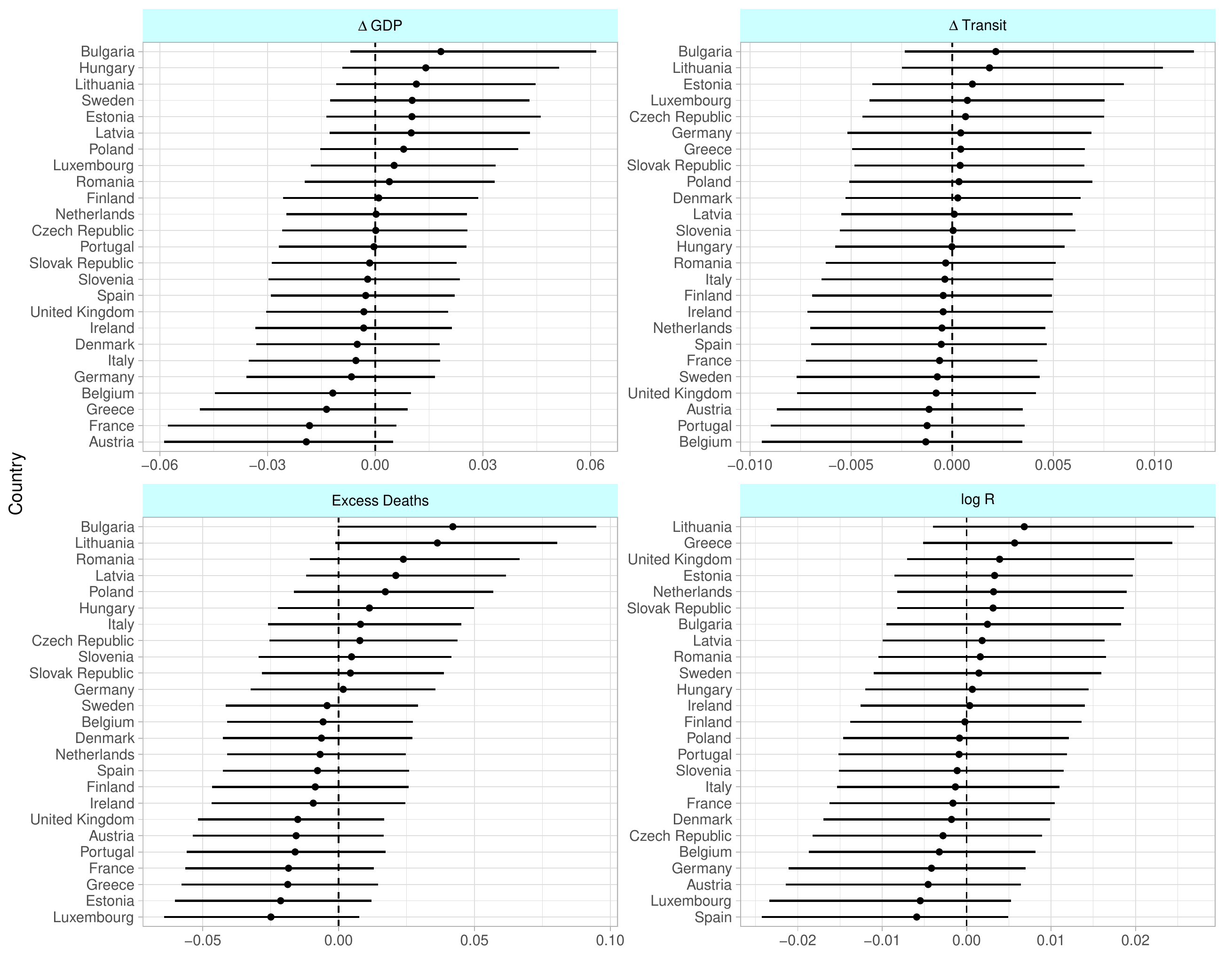}
\caption{Country Effects: Country level effects on response variables (response variable intercept term by country) due to country specific characteristics}\label{random_effects_plot}
\end{figure}

\newpage

\subsection{Country specific characteristics}\label{cnt_spec_covariates}

\begin{table}[!htp]
\centering
\includegraphics[width=1\textwidth]{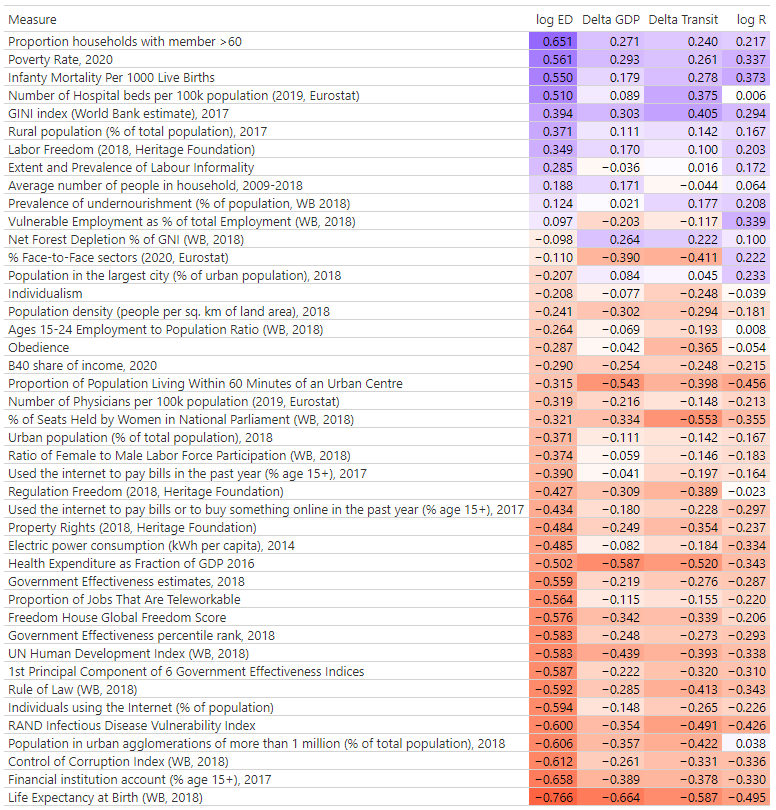}
\caption{Covariates: Correlation of country specific characteristics to country specific response variable intercepts across Government, Societal, economic and Healthcare characteristics.}\label{covariates_result_table}
\end{table}

\begin{figure}[!htp]
\centering
\includegraphics[width=1\textwidth]{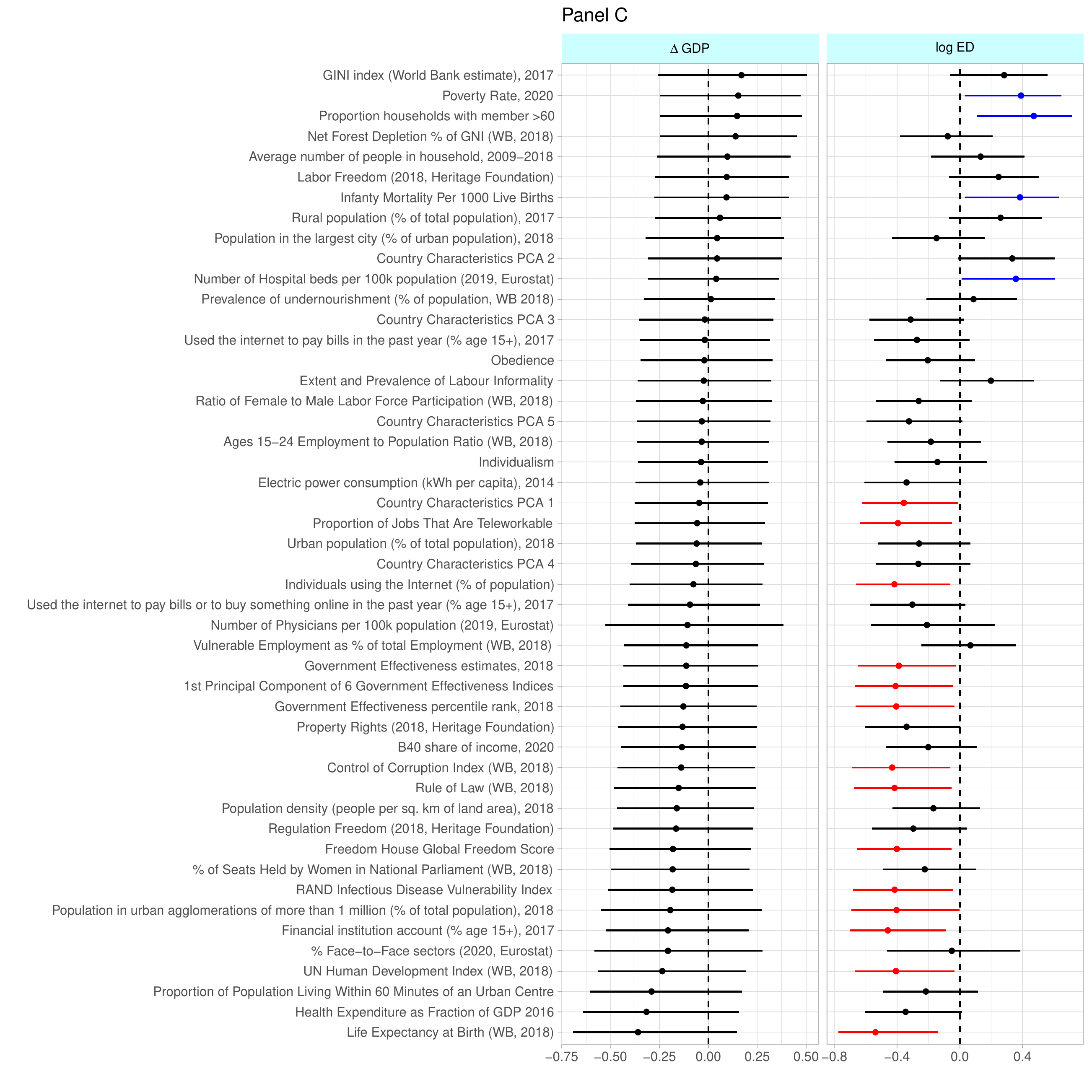}
\caption{Correlations of country specific characteristics with country specific intercepts for $\Delta$ GDP and log ED which are significant at the 90\% CrI. Blue indicates positive and significant correlations and red indicates negative and significant correlations.}\label{covariances_cor_full_1}
\end{figure}

\begin{figure}[!htp]
\centering
\includegraphics[width=1\textwidth]{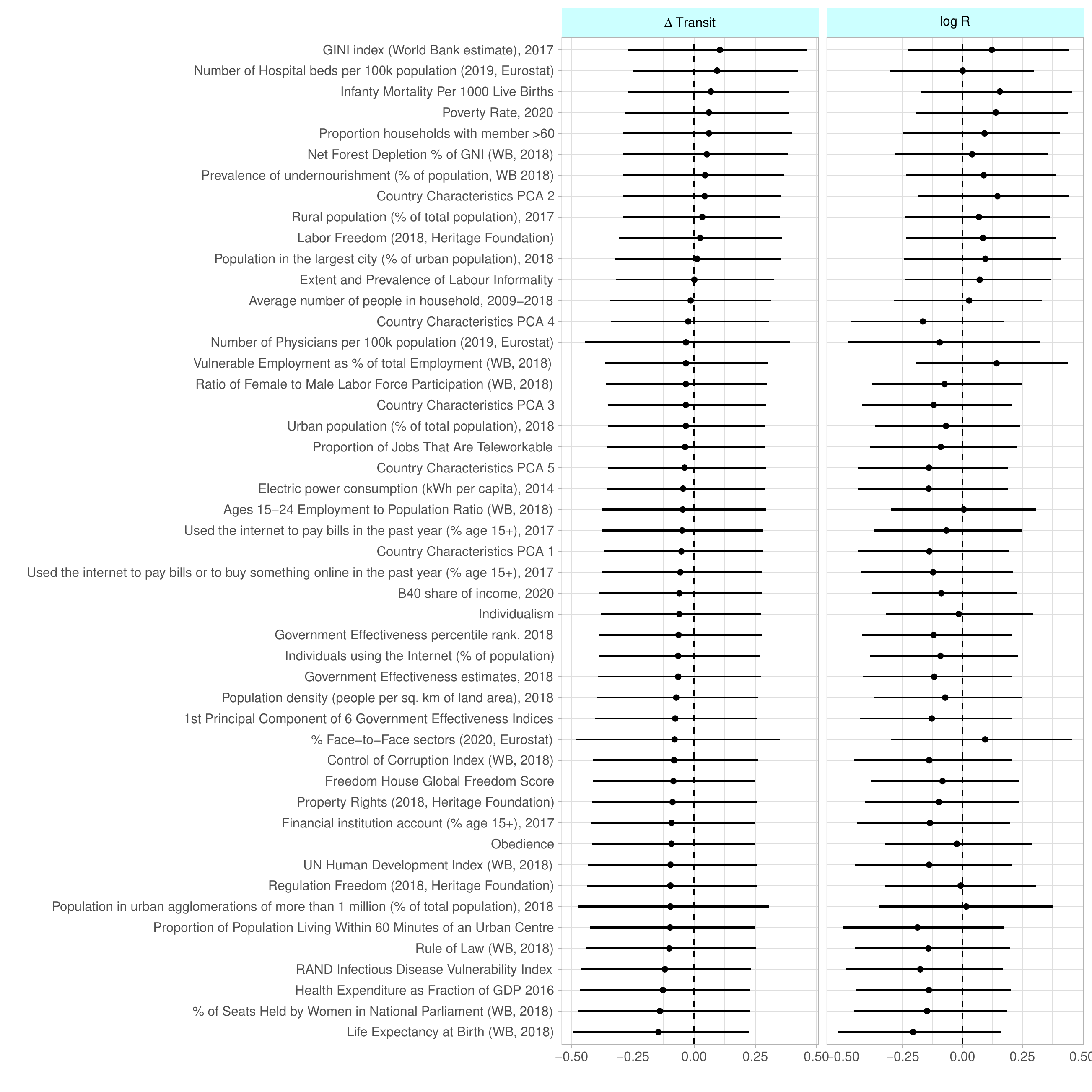}
\caption{Correlations of country specific characteristics
with country specific intercepts for $\Delta$ Transit and log R (none significant at the 90\% CrI).}\label{covariances_cor_full_2}
\end{figure}

\begin{figure}[!htp]
\centering
\includegraphics[width=1\textwidth]{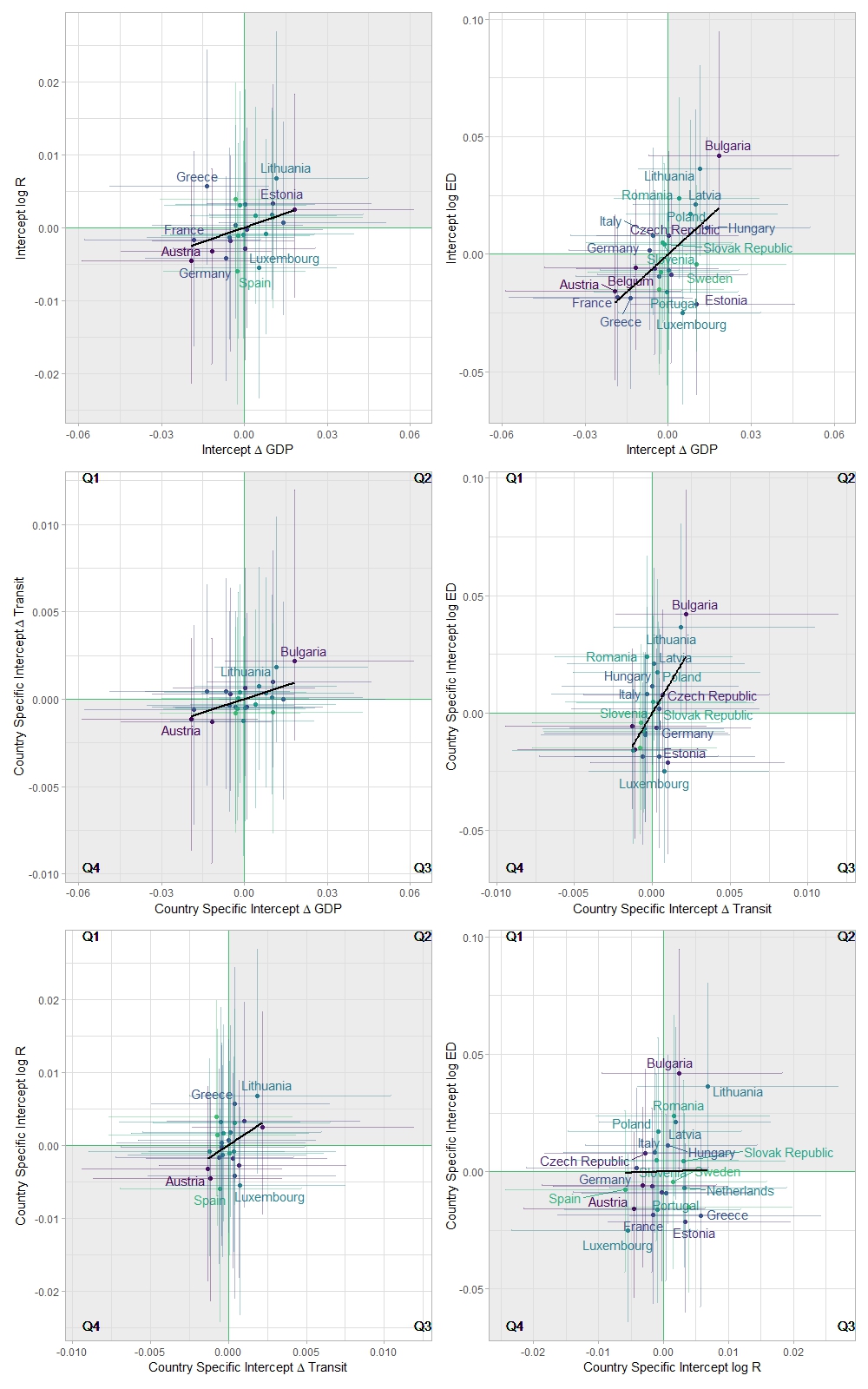}
\caption{Country Effects: Country Level Effects of Response Variables for all Response variable combinations. We highlight the 4 quadrants to display where countries are in the representation.}\label{random_effects_plot_2}
\end{figure}

\subsection{Principal Component Analysis of country specific characteristics}\label{PCA_section}

\begin{figure}[!htp]
\centering
\includegraphics[width=1\textwidth]{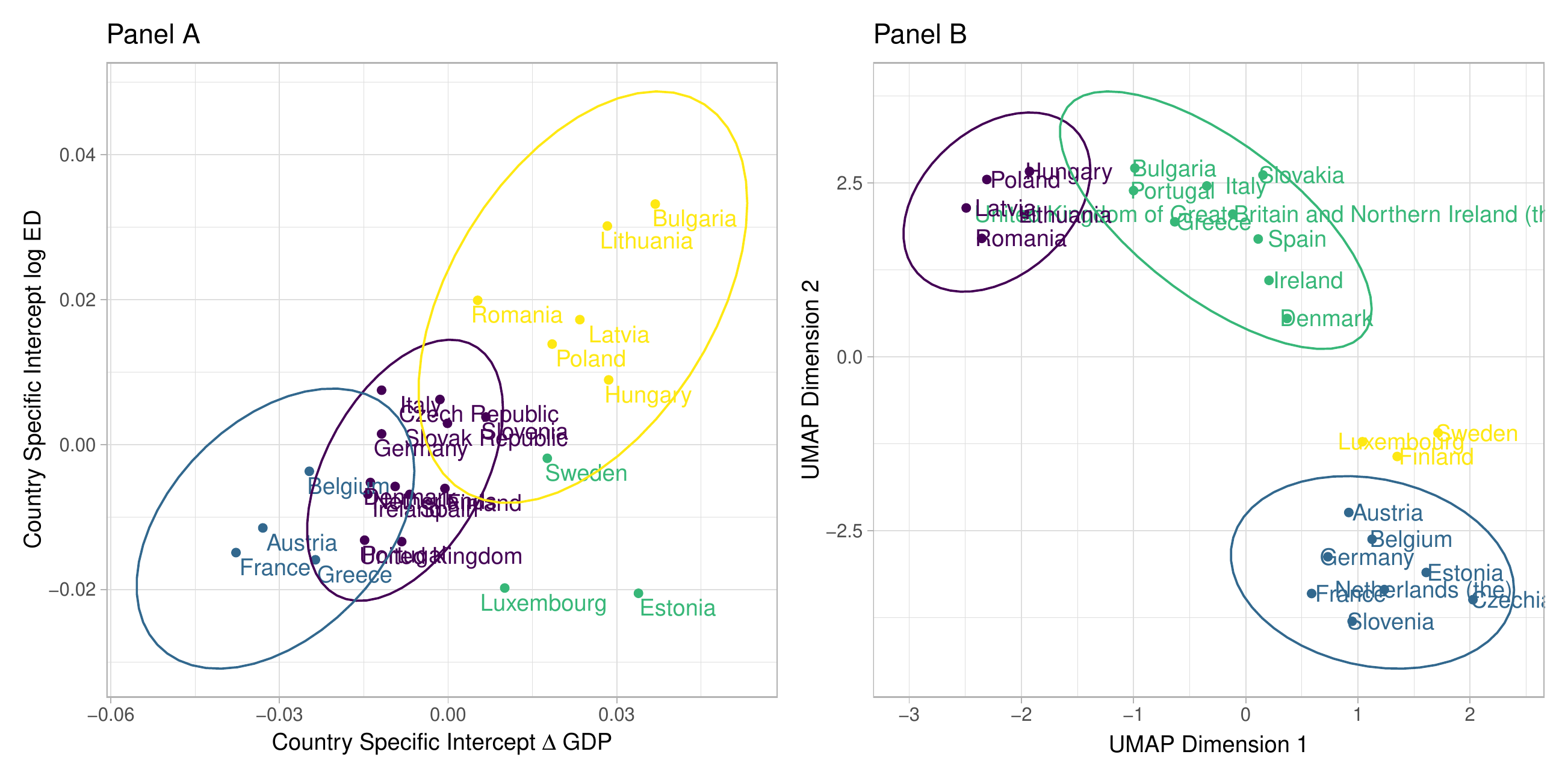}
\caption{(Panel A) Counrty specific intercepts of log ED vs $\Delta$ GDP including k-means country clusters, (Panel B) first 2 dimensions of UMAP decomposition of country specific characteristics including k-means country clusters.}\label{kmeans_plot}
\end{figure}

We can considering different types of clustering by constructing the principal components of the country characteristics, which we provide in Table \ref{pca_table}. The correlation of the principal components is not significant with respect to any of the response variables, which is unsurprising as they are linear combinations of all covariates. We can consider how countries cluster together, either by considering k-means cluster in the plane of country specific intercepts against each other (left hand plot in Figure \ref{kmeans_plot}) or using a dimension reduction technique, here we use UMAP \cite{umap}. The first \& second UMAP components against each other are displayed in the right hand plot of Figure \ref{kmeans_plot}. In both cases we observe that we have clusters which correspond to regional areas in Europe and this is broadly consistent if we just consider country specific characteristics or country specific intercepts, in particular we see clusters around Eastern European countries. 

\singlespacing

\begin{sidewaystable}[!htp]     %
\sidewaystablefn%
\begin{tabular}[t]{lrrrrr}
\toprule
Covariate & Dim.1 & Dim.2 & Dim.3 & Dim.4 & Dim.5\\
\midrule
Government Effectiveness estimates, 2018 & 0.951 & -0.187 & 0.046 & -0.034 & -0.030\\
Control of Corruption Index (WB, 2018) & 0.946 & -0.128 & 0.136 & 0.047 & -0.041\\
Rule of Law (WB, 2018) & 0.918 & -0.217 & 0.154 & -0.006 & -0.057\\
UN Human Development Index (WB, 2018) & 0.914 & 0.108 & 0.124 & -0.048 & 0.070\\
RAND Infectious Disease Vulnerability Index & 0.914 & 0.088 & 0.037 & -0.087 & -0.011\\
Financial institution account (\% age 15+), 2017 & 0.894 & 0.116 & -0.115 & -0.186 & 0.118\\
Used the internet to pay bills or to buy something online ... & 0.893 & -0.013 & 0.170 & -0.051 & 0.048\\
...  in the past year (\% age 15+), 2017 & & & & & \\
Property Rights (2018, Heritage Foundation) & 0.892 & -0.168 & 0.168 & -0.003 & 0.053\\
Individuals using the Internet (\% of population) & 0.867 & -0.133 & 0.231 & 0.048 & 0.064\\
Used the internet to pay bills in the past year (\% age 15+), 2017 & 0.865 & -0.190 & 0.175 & -0.039 & 0.056\\
Proportion of Jobs That Are Teleworkable & 0.853 & 0.017 & -0.052 & 0.123 & 0.061\\
Freedom House Global Freedom Score & 0.836 & -0.110 & -0.074 & -0.049 & 0.181\\
Ratio of Female to Male Labor Force Participation (WB, 2018) & 0.808 & -0.405 & -0.009 & 0.056 & -0.023\\
Regulation Freedom (2018, Heritage Foundation) & 0.761 & -0.231 & 0.053 & 0.048 & 0.190\\
Health Expenditure as Fraction of GDP 2016 & 0.744 & 0.234 & 0.062 & 0.157 & -0.225\\
Electric power consumption (kWh per capita), 2014 & 0.721 & -0.186 & -0.417 & 0.069 & -0.242\\
Life Expectancy at Birth (WB, 2018) & 0.702 & 0.489 & -0.262 & -0.086 & 0.087\\
Urban population (\% of total population), 2018 & 0.632 & 0.188 & -0.121 & 0.679 & 0.032\\
B40 share of income, 2020 & 0.491 & -0.013 & 0.251 & -0.428 & -0.368\\
Number of Physicians per 100k population (2019, Eurostat) & 0.463 & 0.297 & -0.578 & -0.012 & -0.027\\
Population density (people per sq. km of land area), 2018 & 0.357 & 0.577 & 0.459 & 0.381 & -0.152\\
Proportion of Population Living Within 60 Minutes of an Urban Centre & 0.240 & 0.751 & 0.416 & -0.139 & -0.213\\
Obedience & 0.028 & 0.502 & 0.174 & -0.387 & 0.343\\
Labor Freedom (2018, Heritage Foundation) & -0.139 & -0.299 & 0.655 & 0.080 & 0.439\\
Number of Hospital beds per 100k population (2019, Eurostat) & -0.549 & -0.135 & 0.286 & 0.086 & -0.628\\
Extent and Prevalence of Labour Informality & -0.625 & 0.366 & 0.005 & 0.062 & 0.370\\
Rural population (\% of total population), 2017 & -0.632 & -0.188 & 0.121 & -0.679 & -0.032\\
Infanty Mortality Per 1000 Live Births & -0.728 & -0.088 & 0.349 & 0.398 & -0.002\\
Poverty Rate, 2020 & -0.820 & -0.165 & -0.043 & 0.336 & 0.108\\
\botrule
\end{tabular}
\caption{First 5 dimensions (corresponding to the 5 largest eigenvalues) of the Principal Component Analysis of the country specific characteristics}\label{pca_table}%
\end{sidewaystable}

\newpage

\subsection{Forecast comparison}

\begin{figure}[!htp]
\centering
\includegraphics[width=1\textwidth]{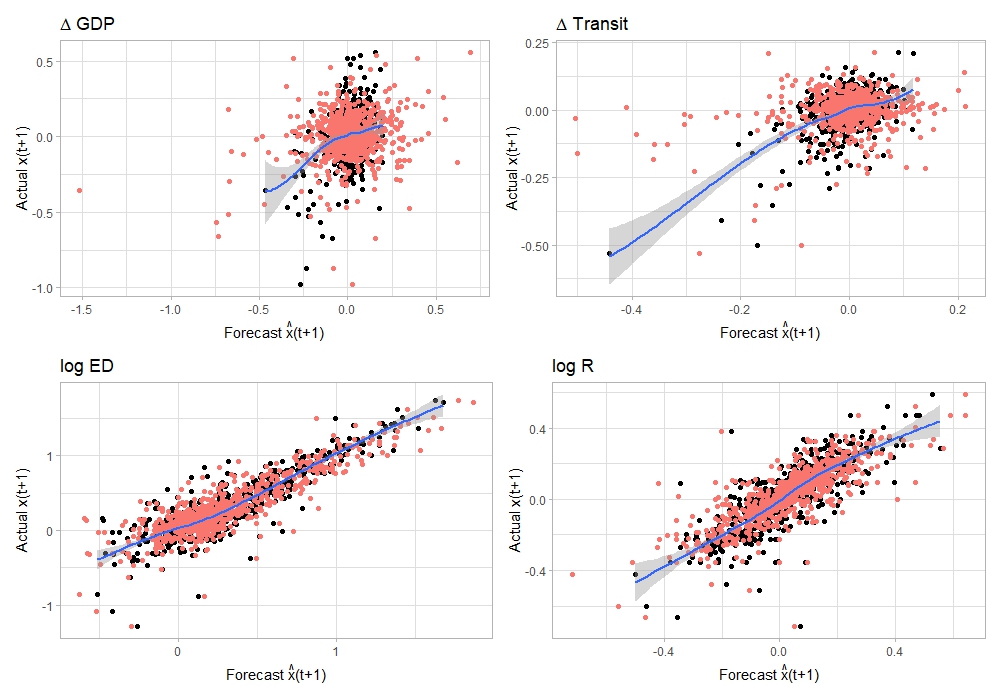}
\caption{Forecast Error: Plot of $\hat{x}(t+1)$ forecast vs actual $x(t+1)$. Black dots are forecast from model, Red dots are naive forecast $\hat{x}(t+1)=x(t)$}\label{forecast_plot}
\end{figure}

\newpage

\subsection{Convergence statistics} \label{convergence_stats}

\begin{figure}[!htp]
\centering
\includegraphics[width=1\textwidth]{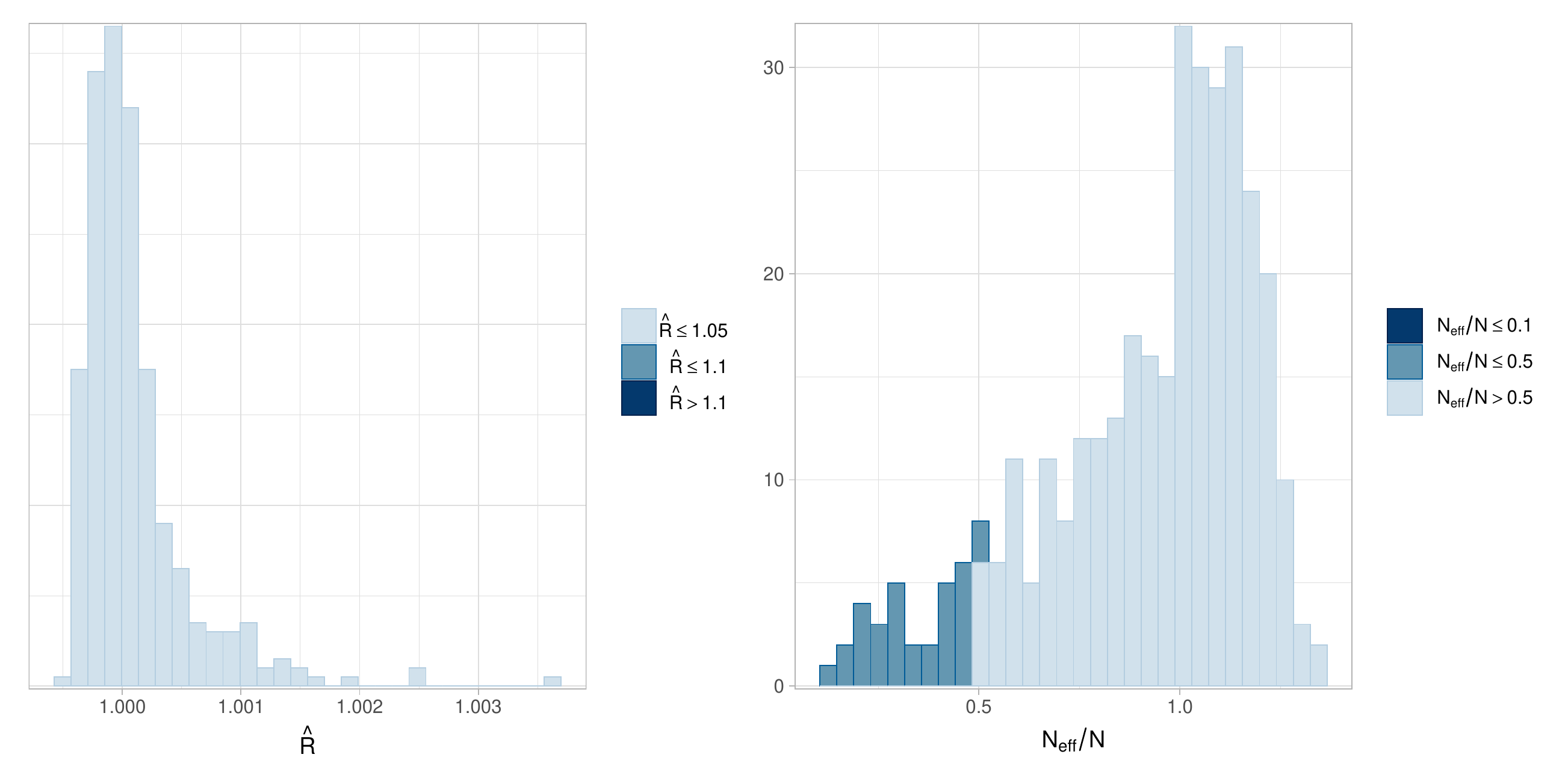}
\caption{MCMC convergence statistics (taken from a run using model values for all parameters as outlined in Table \ref{param_table}). Left: $\hat{R}$ values are close to 1, indicating convergence. Right:
relative Effective sample size .}\label{brms_analytics_plot}
\end{figure}

MCMC convergence statistics are available in Figure \ref{brms_analytics_plot} and show the rank-normalized split $\hat{R}$ score \cite{Vehtari2021-fx} and relative Effective
Sample Size MCMC convergence statistics. Values of $\hat{R}$ close to 1 indicate convergence of the MCMC sampling algorithm. We note the sample properties are such $\hat{R} \leq 1.01$ for all estimated parameters \cite{Vehtari2021-fx} (Figure \ref{brms_analytics_plot}). This implies that the posterior has converged and can be used for inference. 

We used 2000 warmup samples and 2000 iterations per chain, for 4 chains.
We also see that the relative effective sample size exceeds 0.5 for the majority of parameters, indicating low autocorrelation and fast mixing.

\newpage

\subsection{Sensitivity Analysis} \label{sensitivity_analysis}

\begin{figure}[!htp]
\centering
\includegraphics[width=1\textwidth]{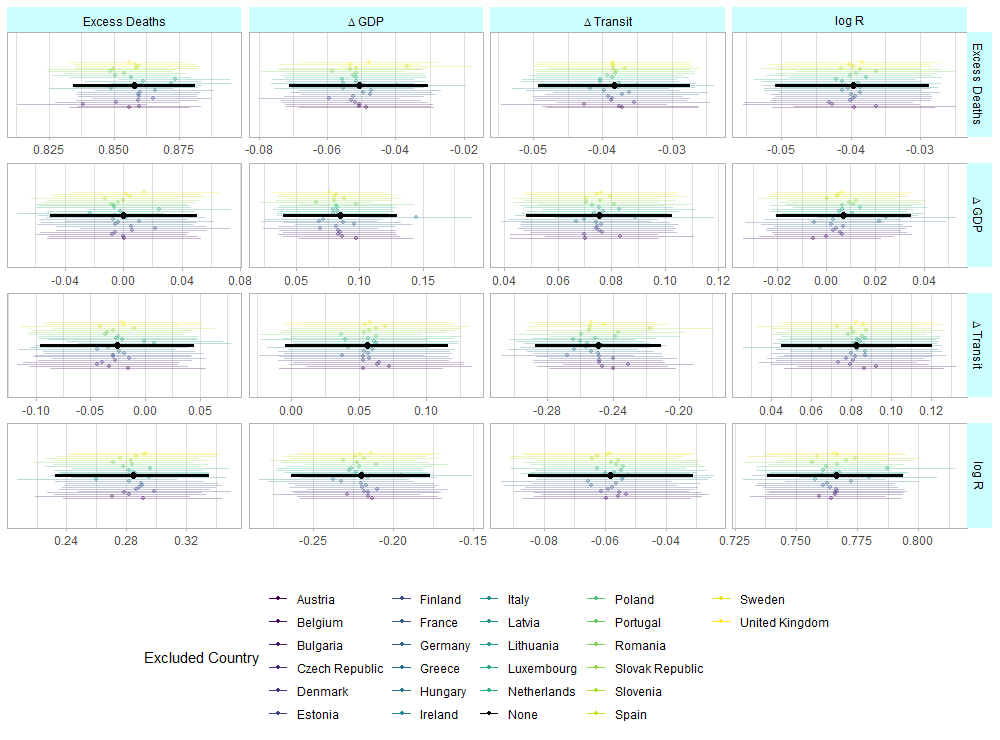}
\caption{Country Sensitivity Analysis: Marginal posterior density 90\% confidence interval and mean for coefficients of the VAR component. Black line is the estimate for the full model, each colour represents the estimate of the model exlcuding that country.}\label{country_sensitivity_plot}
\end{figure}

In order to check that no single country overly influenced the estimation, we re-estimated the model 25 times, each time leaving out one of the countries in the analysis. In Figure \ref{country_sensitivity_plot}, we plot the distribution of coefficients of the VAR component of the model. While there is some variation in the resulting coefficient estimates, this variation is generally small.

We also consider robustness for parameter choices in our model and estimation, such as $\tau$. The variable selection broadly follows the approach taken in  \cite{liu2021}.  

There are a range of different specification for the model. Rather than estimating the model with no intercept we could introduce a global intercept and impose a constraint on the wild type variant coefficient to equal zero. The estimated coefficients are essentially unchanged by this.  

\end{document}